\documentclass[fleqn,usenatbib]{mnras}

\usepackage{newtxtext,newtxmath}
\usepackage[T1]{fontenc}
\usepackage{ae,aecompl}

\usepackage{graphicx}	% Including figure files
\usepackage{amsmath}	% Advanced maths commands
\usepackage{amssymb}	% Extra maths symbols
\usepackage{threeparttable}

\defcitealias{DL17}{Dierickx \& Loeb(2017)}
\providecommand{\dt}[1]{{\tt #1}} % for field names in the data model

\title[The Substructures of the Milky Way]{Probing the Galactic Halo with RR Lyrae Stars. II. The Substructures of the Milky Way}

\author[Wang et al.]{
F. Wang$^{1,2}$\thanks{E-mail: fwang\_kiaa@pku.edu.cn (FW); zhanghw@pku.edu.cn (HWZ)},
H.-W. Zhang$^{1,2}$\footnotemark[1],
X.-X. Xue$^{3}$,
Y. Huang$^{4}$,
G.-C. Liu $^{5}$,
L. Zhang$^{3}$,
and 
\newauthor
C.-Q. Yang$^{6}$
\\
$^{1}$Department of Astronomy,  School of Physics, Peking University, Beijing 100871, P.R. China.\\
$^{2}$Kavli Institute for Astronomy and Astrophysics,
Peking University,
Beijing 100871, P.R. China.\\
$^{3}$CAS Key Laboratory of Optical Astronomy, National Astronomical Observatories, Chinese Academy of Sciences, 20A Datun Road, Chaoyang District, \\
Beijing 100101, P.R. China. \\
$^{4}$South-Western Institute for Astronomy Research,
Yunnan University,
Kunming, Yunnan 650091, P.R. China. \\
$^{5}$Center for Astronomy and Space Sciences, China Three Gorges University, Yichang 443002, P.R. China.\\
$^{6}$Shanghai Astronomical Observatory, 80 Nandan Road, Shanghai 200030, P.R. China.}

\date{Accepted 2022 March 28. Received 2022 March 27; in original form 2021 November 30}

\pubyear{2022}

\begin{document}

\maketitle
\begin{abstract}
We identify substructures of the Galactic halo using 3,003 type $ab$ RR Lyraes (RRab) with 6D position-velocity information from the SDSS, LAMOST, and Gaia EDR3. Based on the information, we define the separation of any two of the stars in the integrals of motion space and identify substructures by utilizing the friends-of-friends algorithm. We identify members belonging to several known substructures: the Sagittarius stream, the Gaia-Enceladus-Sausage (GES), the Sequoia, and the Helmi streams. In addition to these known substructures, there are three other substructures possibly associated with globular clusters NGC 5272, NGC 6656, and NGC 5024, respectively. Finally, we also find three remaining unknown substructures and one of them has large angular momentum and a mean metallicity $\rm -2.13\,dex$ which may be a new substructure.
As for GES, we find that it accounts for a large part of substructures in the inner halo and the range of apocenter distance is from 10 to $34\,\rm kpc$, which suggests that the GES is mainly distributed in the inner halo. The near one-third proportion of the GES and the peak value $20\,\rm kpc$ of the apocenter distances suggest that GES could account for the break in the density profile of the Galactic halo at Galactocentric distance ${\sim}20-25\,\rm kpc$. The similarity of comparing the kinematic properties of Gaia-Enceladus-Sausage with the Hercules-Aquila Cloud and Virgo Overdensity suggests that the three substructures may have similar origins.

\end{abstract}


%%%%%%%%%%%%%%%%%%%%%%%%%%%%%%%%%%%%%%%%%%%%%%%%%%

%%%%%%%%%%%%%%%%% BODY OF PAPER %%%%%%%%%%%%%%%%%%

\section{Introduction} \label{sec:intro}

According to the $\rm {\Lambda}CDM$ cosmological model, the large-scale structure forms through hierarchical processes (Peebles 1974; White \& Rees 1978; Blumenthal et al. 1984). The hierarchical structure formation model implies that the formation of our Milky Way (MW) involves a series of accretion and merger events (e.g., Searle \& Zinn 1978; White \& Rees 1978; Blumenthal et al. 1984). The accreted satellite galaxies would be tidally disrupted, leaving stellar debris and resulting in substructures in the stellar halo (e.g., Bullock et al. 2001; Bullock \& Johnston 2005; Cooper et al. 2010). In a gravitationally collisionless system, the phase-space distribution of stars would be preserved and could be the evidence of their different origins, which suggests that identifying the substructure in the phase space is an important tool to study the formation history of the MW. Moreover, chemical properties could be used to further constrain the origins of those substructures (Freeman \& Bland-Hawthorn 2002).

In the past decades, substructures have been identified using photometric surveys, e.g. Sloan Digital Sky Survey (SDSS; York et al. 2000), Two Micron All Sky Survey (2MASS; Skrutskie et al. 2006), and PanSTARRS1 (PS1; Chambers et al. 2016). The most prominent and important discoveries are the Sagittarius dwarf galaxy (Ibata et al. 1994, 1995; Yanny et al. 2000) and its tidal streams (Ibata et al. 2001; Majewski et al. 2003). With the development of the spectroscopic surveys and astrometric surveys, the 6D information of positions and velocities can be measured for the sample in the solar vicinity and substructures can be identified in the position-velocity space or even in the integrals of motion (IoM) space (Helmi et al. 1999; Klement et al. 2008, 2009; Morrison et al. 2009; Smith et al. 2009). The accurate position, parallax, and proper motions for over one billion stars released in Gaia DR2 (Gaia Collaboration et al. 2016, 2018) provide a good opportunity to investigate substructures in the local Galactic halo. For example, `Gaia Sausage' or `Gaia-Enceladus', a massive merger event, has been identified in the inner halo (Belokurov et al. 2018; Haywood et al. 2018; Helmi et al. 2018; Myeong et al. 2018). Different from the Gaia Sausage, the Sequoia, another merger event, is proposed and the merger debris is in the retrograde motions and more metal-poor (Myeong et al. 2019). But for the distant halo, the distances estimated by the parallax measurements could not be used due to the large uncertainties of parallax. In this case, some tracers, e.g. K giants, blue horizontal-branch (BHB), and RR Lyrae (RRL) stars, have been used to identify substructures in the distant halo (Yang et al. 2019a; Yuan et al. 2020; Helmi 2020).

RR Lyrae stars (RRLs) are old and metal-poor variable stars with a well-defined luminosity-metallicity relation in the optical band and period-metallicity-luminosity (PMZ) relation in the infrared band, which makes them good standard candles. These properties indicate that RRLs are ideal tracers to study the Galactic halo (Zinn et al. 2004; Keller et al. 2008; Sesar et al. 2013; Sesar et al. 2017b). But due to the pulsation of RRLs, the radial velocities will vary with the pulsation phase. Liu et al. (2020) measured the systemic radial velocities and metallicities for thousands of RRLs by using large-scale spectroscopic surveys, SDSS/Sloan Extension for Galactic Understanding and Exploration (SEGUE; Yanny et al. 2009) and the Large Sky Area Multi-Object Fiber Spectroscopic Telescope (LAMOST; Cui et al. 2012; Deng et al. 2012; Zhao et al. 2012; Liu et al. 2014). Combining their catalogue with proper motions from Gaia Early Data Release 3 (Gaia EDR3; Gaia Collaboration et al. 2020), we obtain a large RRL sample with 6D position-velocity information and even metallicities. The precise distances with median uncertainties of about 7\% allow us to identify substructures in the IoM space. Besides, we also increase the number of RRLs by combining the recently published RRL catalogues from photometric surveys with more spectra from LAMOST and SDSS. Using this large RRL sample, we aim at identifying substructures and studying their properties.

This paper is organized as follows. The data employed in this work is described in Section 2 and the group identification approach is shown in Section 3. We present the results in Section 4. Finally, a brief summary is presented in Section 5.

\section{Data} \label{sec:style}

\begin{figure}
	\centering
	\includegraphics[width = \linewidth]{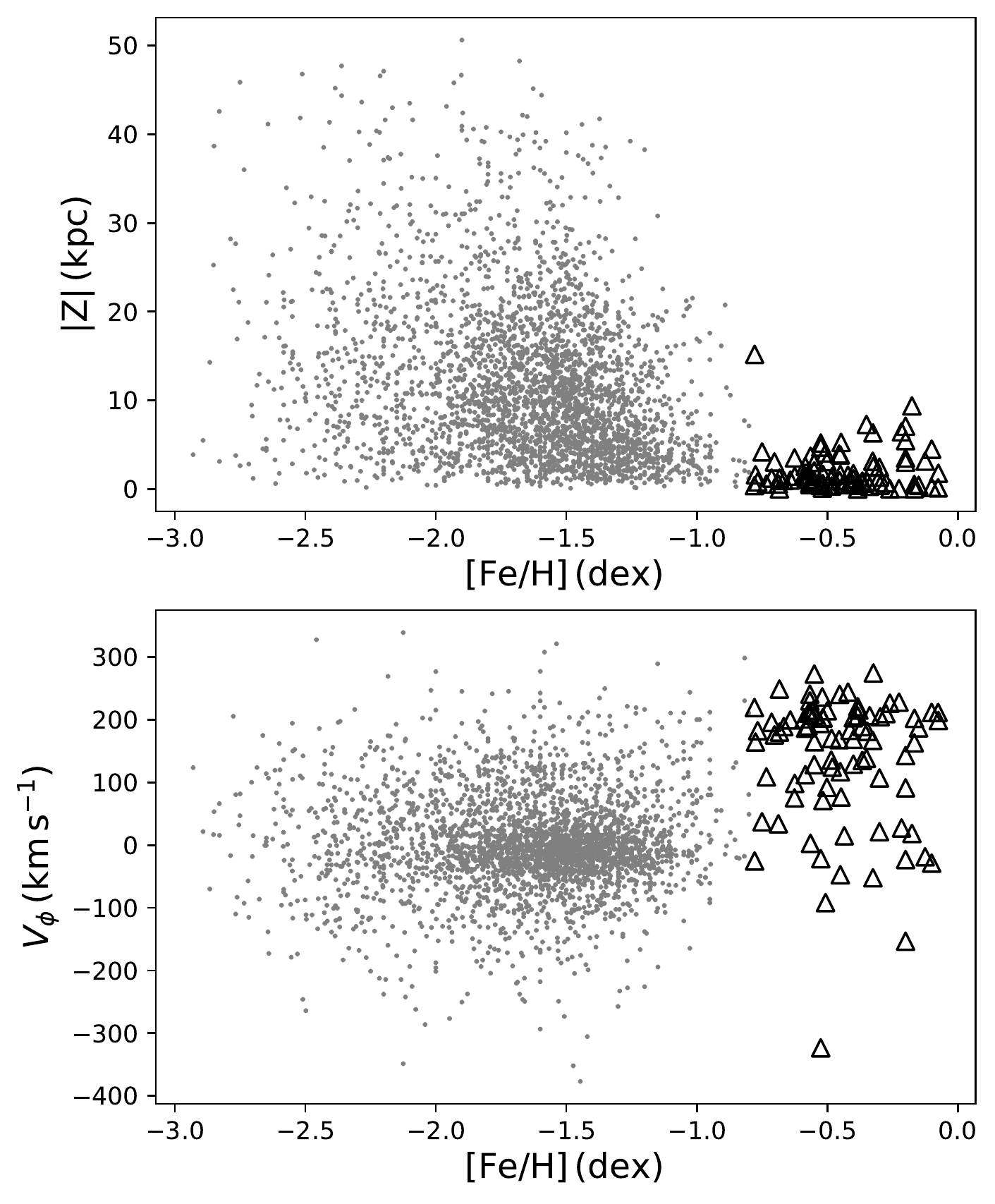}
	\caption{
	The distributions of relatively metal-poor ($\rm [Fe/H]<-0.8\,\rm dex$) and metal-rich ($\rm [Fe/H]>-0.8\,\rm dex$) RRab in the $({\rm [Fe/H]},|Z|)$ and $({\rm [Fe/H]},V_\phi)$ space. The gray dots and the black triangles represent the metal-poor and metal-rich RRab, respectively.}
	\label{fig:1}
\end{figure}

\begin{figure*}
	\centering
	\includegraphics[width = \linewidth]{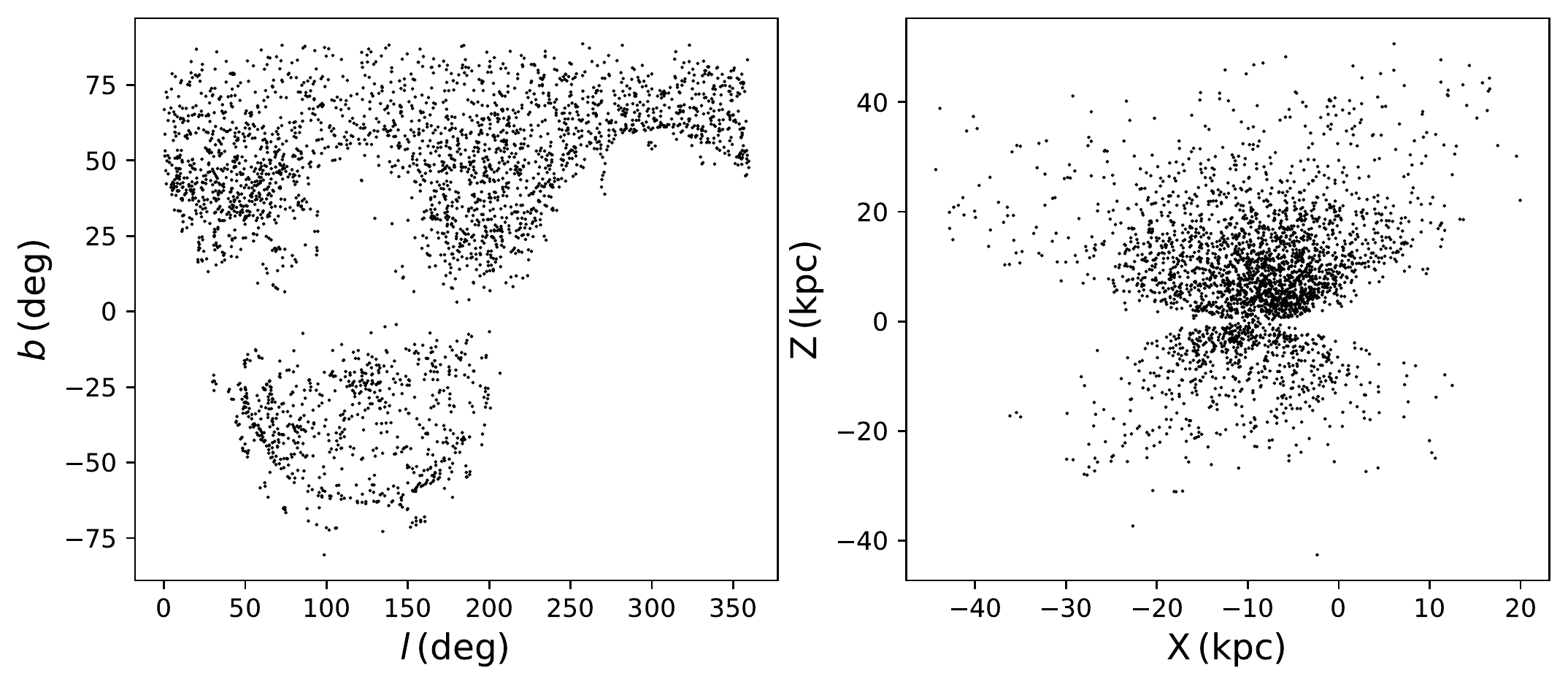}
	\caption{The distribution in the Galactic coordinate system $(l,b)$ and the spatial distribution in the $X-Z$ plane of our halo sample.}
	\centering
	\label{fig:2}
\end{figure*}

\begin{figure*}
	\centering
	\includegraphics[width = \linewidth]{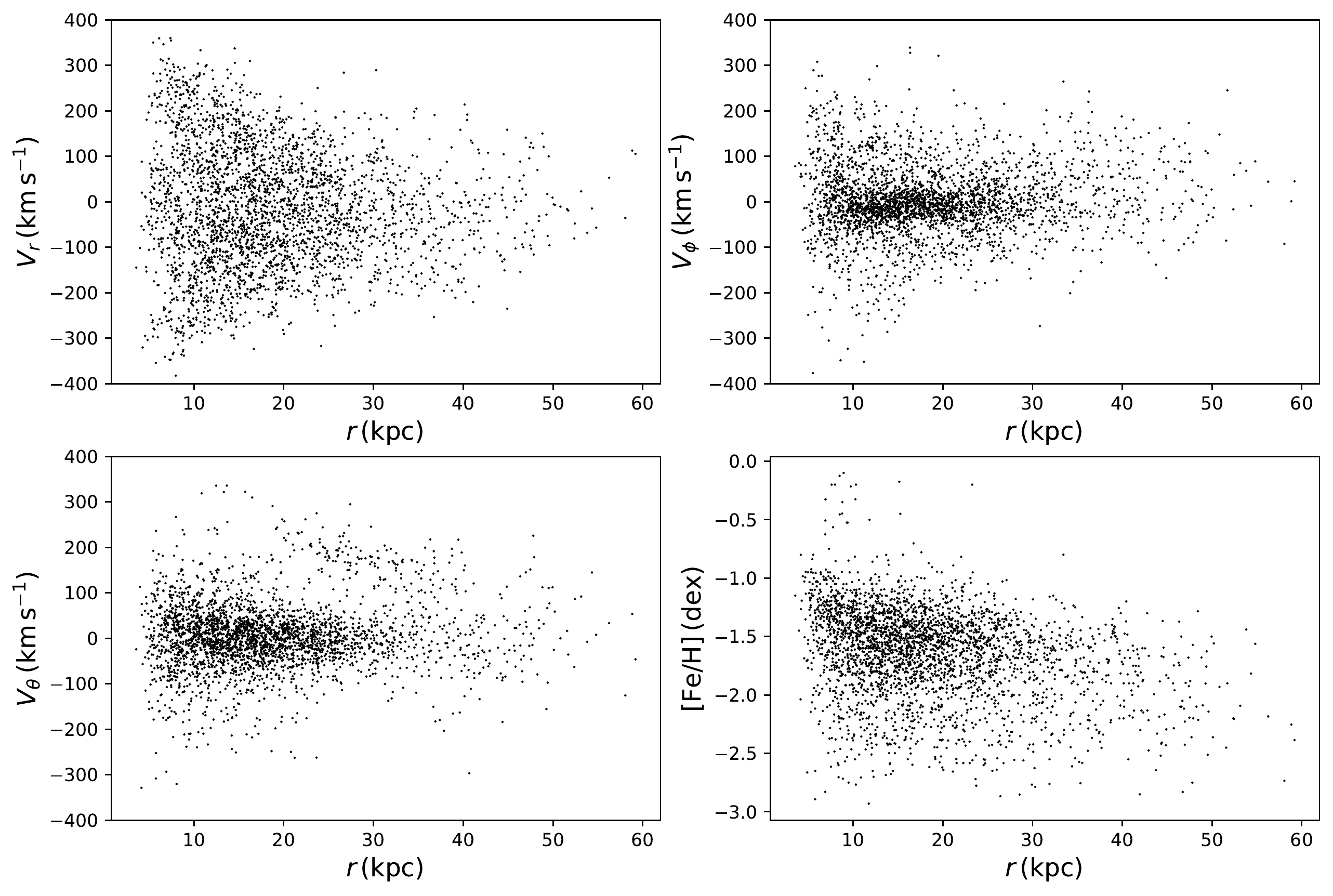}
	\caption{The velocity and metallicity distributions along with the Galactocentric distance $r$ of our halo sample.}
	\centering
	\label{fig:3}
\end{figure*}

\subsection{Coordinate systems}
We use a right-hand Galactocentric Cartesian coordinate system $(X,Y,Z)$, a Galactocentric cylindrical coordinate system $(R,\phi,Z)$ and a Galactocentric spherical coordinate system $(r,\theta,\phi)$. For the Cartesian coordinate system, $X$ points in the direction opposite to the Sun, $Y$ is in the direction of Galactic rotation and $Z$ is towards the North Galactic Pole. For the cylindrical coordinate system, $R$ is the projected Galactocentric distance and $Z$ is the same as that in the Cartesian system. $\phi$ is the azimuthal angle between the direction from the Galactic center towards the Sun and the direction to the projected position of the star. For the spherical coordinate system, $r$ is the Galactocentric distance, $\theta$ increases from 0 to $\pi$ from the North Galactic Pole to the South Galactic Pole and $\phi$ is the same as that in the cylindrical coordinate system. The three velocity components are represented by $(U,V,W)$, $(V_R,V_\phi,V_z)$ and $(V_r,V_\theta,V_\phi)$ corresponding to the Cartesian, cylindrical and spherical coordinate system, respectively. The Sun's position is at $(X,Y,Z)=(-R_0, 0, 0)\,\rm kpc$ which $R_0$, the distance from the Sun to the Galactic center, is $8.0\,\rm kpc$ (Reid 1993). We adopt the solar peculiar velocity of $(11.69, 10.16, 7.67)\,\rm km\,s^{-1}$ (Wang et al. 2021) and the local standard of rest (LSR) velocity is $220\,\rm km\,s^{-1}$ (Kerr \& Lynden-Bell 1986).

\subsection{RRL sample}
In this work, the RRL catalogue we have adopted is from Liu et al. (2020). They publicated a catalogue of 5290 RRLs with metallicities estimated from spectra of the LAMOST (Deng et al. 2012; Zhao et al. 2012) and SDSS/SEGUE surveys (Yanny et al. 2009). They also estimated the systemic radial velocities $V_{\rm los}$ for 3,642 objects by fitting empirical templates to the velocity curves of the multiple measurements. The typical precision of metallicities is $0.2\,\rm dex$ and the uncertainties of the systemic radial velocities are in the range of 5 to $21\,\rm km\,s^{-1}$ which depend on the number of radial-velocity measurements available for a specific star. The distance estimates are from multi-band photometries with median uncertainties of about 7\%. In this catalogue, we just select 2,246 type $ab$ RRLs (RRab) with metallicities, systemic radial velocities, and distance estimates, because the estimates of the distance and radial velocity for RRab are more precise.

In addition to the catalogue from Liu et al. (2020), we enlarge the dataset by combining the recently published RRL catalogue from photometric surveys with the spectra from LAMOST and SDSS. Besides the RRL photometric catalogue from Liu et al. (2020), we collect the RRL catalogue from Gaia (Clementini et al. 2019), the All-Sky Automated Survey for SuperNovae (ASAS-SN, Shappee et al. 2014; Jayasinghe et al. 2019) and PS1 (Sesar et al. 2017a). To ensure the purity of the RRL sample, we just select the RRLs with classification scores larger than 0.8 in the PS1 catalogue and classification probabilities larger than 0.8 in the ASAS-SN catalogue. We also utilize the newest RRL catalogues from the Catalina Survey (Drake et al. 2013a, 2013b, 2014, 2017; Torrelaba et al. 2015) and General Catalogue of Variable Stars (GCVS; Samus' et al. 2017). For the spectroscopic data set, we utilize the spectra from SDSS Data Release 12 (SDSS DR12; Alam et al. 2015) and a larger quantity of spectra from LAMOST DR6. We adopt the same method as Liu et al. (2020) to estimate the systemic radial velocities, metallicities, and distances. We measure the radial velocity of each single-exposure spectrum from three Balmer lines, $\rm H\alpha$, $\rm H\beta$, and $\rm H\gamma$, respectively, and we estimate the systemic radial velocities utilising the radial velocity curve template from Sesar et al. (2012) for these three Balmer lines. The metallicities are estimated by the template-matching method using a series of synthetic spectra. As for the distances, we use the relation between absolute visual magnitude and metallicity and PMZ relation in near- or mid-infrared magnitude (see Liu et al. 2020, for more details). The final total number of our RRab sample is 3,193.

By cross-matching with the Gaia EDR3 catalogue (Gaia Collaboration et al. 2021), we get proper motions for these sample stars. We select the Gaia EDR3 data with \dt{ruwe}\,$<1.4$ to remove stars with dubious astrometry (Fabricius et al. 2021). Using the python package \texttt{galpy} (Bovy 2015), we calculate the Galactic longitude $l$, latitude $b$, and tangential velocities ($V_l,V_b$) based on the right ascension R.A., declination Dec., proper motions, and distances. With the solar peculiar velocity and the LSR velocity, all velocities of stars are converted to the Galactic standard of rest (GSR) frame. We exclude the stars with the uncertainty $\sigma$ of any velocity component larger than $100\,\rm km\,s^{-1}$. Then we remove a few stars with the very large total velocity, $V_{\rm tot}=\sqrt{{V_{\rm los}}^2+{V_l}^2+{V_b}^2}>400\,\rm km\,s^{-1}$, or with the large semimajor axis $a$, because the parameters of these stars may be unreliable. Finally, we get a sample of 3,065 RRab stars with full 6D information (3D positions and 3D velocities). Table\,1 shows an example for the parameters and corresponding uncertainties of our sample.

\begin{table*}
	\centering
	\caption{The parameters of our all 3,065 RRab samples.}
	\label{tab:1}
	\resizebox{2\columnwidth}{!}{
	\begin{tabular}{cccccccccccccccc} % four columns, alignment for each
		\hline
index & ra & dec & $d$ & $d_{\rm err}$ & $V_{\rm los}$ & $V_{\rm los,err}$ & pmra & $\rm pmra_{err}$ & pmdec & $\rm pmdec_{err}$ & [Fe/H] & $\rm [Fe/H]_{err}$ & period & source\\
& $\rm (deg)$ & $\rm (deg)$ & $(\rm kpc)$ & $(\rm kpc)$ & $(\rm km\,s^{-1})$ & $(\rm km\,s^{-1})$ & $(\rm mas\,yr^{-1})$ & $(\rm mas\,yr^{-1})$ & $(\rm mas\,yr^{-1})$ & $(\rm mas\,yr^{-1})$ & $(\rm dex)$ & $(\rm dex)$ & $(\rm day)$ \\
\hline

0 & 267.264465 & 29.84226 & 14.599 & 1.157 & $-164.626$ & 5.698 & $-2.469$ & 0.042 & $-3.468$ & 0.049 & $-1.435$ & 0.164 & 0.579167 & Gaia \\
1 & 269.346801 & 9.682361 & 5.062 & 0.25 & 33.6 & 5.246 & $-2.108$ & 0.024 & $-3.773$ & 0.02 & $-1.271$ & 0.106 & 0.718899 & Gaia \\
2 & 279.531531 & 7.919324 & 5.842 & 0.452 & 30.171 & 5.698 & $-0.818$ & 0.041 & $-6.152$ & 0.036 & $-0.367$ & 0.195 & 0.471039 & ASASSN \\
3 & 297.675479 & 39.480081 & 2.643 & 0.13 & $-236.572$ & 5.246 & $-12.539$ & 0.013 & $-27.911$ & 0.014 & $-1.688$ & 0.124 & 0.550249 & Gaia \\
4 & 318.98199 & 14.948555 & 8.148 & 0.536 & $-24.871$ & 5.698 & $-1.388$ & 0.037 & $-1.462$ & 0.026 & $-1.051$ & 0.11 & 0.630087 & Gaia \\
5 & 322.53859 & 12.226427 & 10.026 & 0.841 & $-141.288$ & 5.246 & $-0.647$ & 0.04 & $-4.009$ & 0.031 & $-2.166$ & 0.106 & 0.574965 & PS1 \\
6 & 339.22582 & 28.042272 & 14.556 & 0.984 & $-130.824$ & 3.179 & 1.112 & 0.083 & 0.134 & 0.087 & $-1.352$ & 0.188 & 0.611899 & Gaia \\
7 & 355.90062 & 45.205608 & 16.36 & 1.108 & $-91.581$ & 3.179 & 0.912 & 0.05 & $-0.615$ & 0.044 & $-1.75$ & 0.398 & 0.601001 & Gaia \\
8 & 205.59049 & 28.425794 & 9.866 & 0.531 & $-153.099$ & 5.484 & $-0.131$ & 0.04 & $-2.485$ & 0.023 & $-1.632$ & 0.105 & 0.559101 & Gaia \\
9 & 302.60155 & $-12.939593$ & 24.681 & 1.955 & $-75.465$ & 5.484 & 0.116 & 0.142 & $-1.322$ & 0.091 & $-1.946$ & 0.082 & 0.643254 & Gaia \\
		\hline
		\multicolumn{10}{l}{\small (This table is available in its entirety in machine-readable form.)} \\
	\end{tabular}}
\end{table*}

\subsection{Halo star selection}
RRLs are old and metal-poor stars and most of them are distributed in the Galactic halo. However, some relatively metal-rich RRLs have similar kinematic and chemical properties to the Galactic disk (Layden et al. 1996; Marsakov et al. 2018; Prudil et al. 2020). Chadid et al. (2017) found that their RRL sample could be divided into metal-rich (disk) and metal-poor (halo) groups at $\rm [Fe/H]=-1.0\,\rm dex$. Luo et al. (2021) also found a significant gap at $\rm [Fe/H]=-0.5\,dex$ in the $\rm [Fe/H]$-period space, which can be used to separate the RRab stars into two groups. As shown in Fig.\,1, based on different distributions in the vertical distances and the azimuthal velocities, our sample can also be divided into the metal-rich (disk) and metal-poor (halo) components at $\rm [Fe/H]\sim-0.8\,\rm dex$. So in order to select the halo RRab as much as possible and reduce the impact of the disk RRab, we exclude 62 RRab with $\rm [Fe/H]>-0.8\,dex$ and $|Z|<3\,\rm kpc$, which occupies the majority of the disk RRab sample. Our final halo RRab sample is 3,003 and the spatial distribution of these stars in the Galactic coordinate system and $X-Z$ plane are shown in Fig.\,2. The distributions of velocities and metallicities are shown in Fig.\,3.

\begin{table}
	\centering
	\caption{The maximum physical sizes of each component corresponding to the linking length of 0.150 and 0.304.}
	\label{tab:2}
	\begin{tabular}{lcccr} % four columns, alignment for each
		\hline
		Linking Length & $\theta(l_{\rm orbit},b_{\rm orbit})$ & $a$ & $e$ & $l_{\rm apo}$\\
		 & $\,\rm (deg)$ & $\,\rm (kpc)$ & & $\,\rm (deg)$\\
		\hline
		0.150 & 14.70 & 2.72 & 0.04 & 15.54\\
		0.304 & 29.80 & 5.51 & 0.09 & 31.49\\
		\hline
	\end{tabular}
\end{table}

\begin{table*}
	\centering
	\caption{The properties of the known substructures in our sample.}
	\label{tab:3}
	\setlength{\tabcolsep}{4mm}{
	\begin{tabular}{lccccc} % four columns, alignment for each
		\hline
		Substructure & Number & $r\,\rm (kpc)$ & $\rm [Fe/H]\,(dex)$ &  $a\,\rm (kpc)$ & $e$\\
		\hline
        Sgr leading arm & 108 & $[29.88, 6.13]$ & $[-1.71, 0.29]$ & $[29.90, 5.72]$ & $[0.38, 0.13]$\\
        Sgr trailing arm & 37 & $[29.44, 5.41]$ & $[-1.69, 0.24]$ & $[55.74, 8.83]$ & $[0.68, 0.07]$\\
        GES & 1067 & $[16.04, 5.84]$ & $[-1.56, 0.31]$ & $[10.84, 3.42]$ & $[0.86, 0.09]$\\
        Sequoia & 99 &  $[22.22, 6.05]$ & $[-1.69, 0.32]$ & $[17.20, 4.45]$ & $[0.66, 0.12]$\\
        Helmi streams & 32 &  $[13.19, 2.91]$ & $[-1.54, 0.45]$ & $[11.61, 2.08]$ & $[0.40, 0.08]$\\        
		\hline
    \multicolumn{4}{l}{\small \textbf{Note}: The values in the brackets represent the mean and the standard deviation.} \\
	\end{tabular}}
\end{table*}

\begin{figure*}
	\centering
	\includegraphics[width = \linewidth]{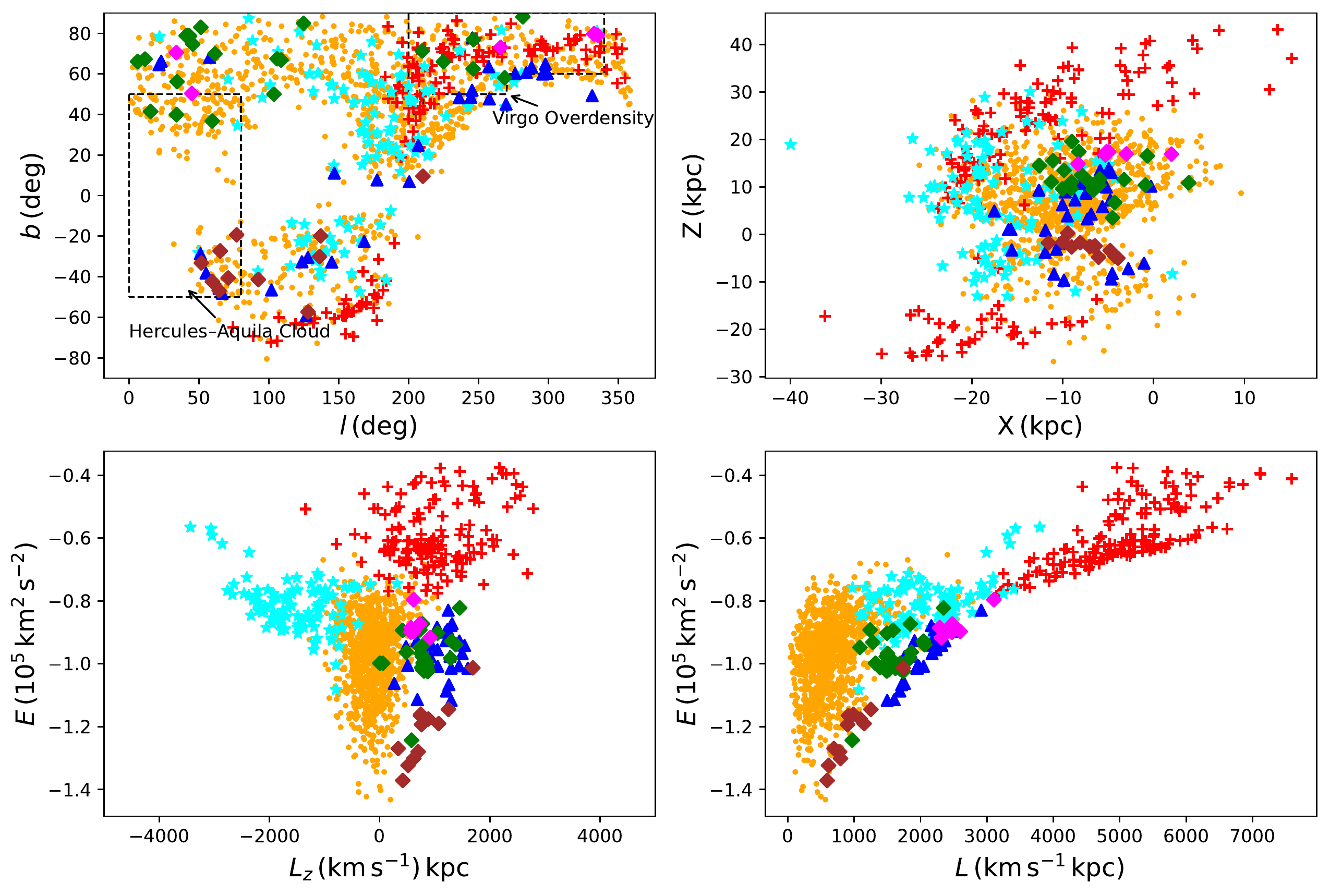}
	\caption{The distributions of the groups associated with known substructures or GCs in the $(l,b)$, $(X,Z)$, $(L_z,E)$, and $(L,E)$ space, respectively. The different colors and symbols represent different substructures: GES (orange dots), Sgr stream (red pluses), Sequoia (cyan stars), and the Helmi streams (blue triangles). The green, brown, and magenta diamond symbols represent the stars possibly associated with the GCs, NGC 5272, NGC 6656, and NGC 5024, respectively. The black dashed areas in the upper left panel represent the regions of the Hercules-Aquila Cloud (Belokurov et al. 2007) and Virgo Overdensity (Newberg et al. 2002).}
	\centering
	\label{fig:4}
\end{figure*}

\section{Method} %\label{sec:style}
In this work, we aim at identifying substructures in the IoM space. Xue et al. (2022, in preparation) defined five IoM parameters: eccentricity $e$, semimajor axis $a$, direction of the orbital pole $(l_{\rm orbit},b_{\rm orbit})$ and the angle between apocenter and the projection of $X$-axis on the orbital plane $l_{\rm apo}$. We use a similar method to the one described in Starkenburg et al. (2009) to define the separation between two stars in $(e,a,l_{\rm orbit},b_{\rm orbit}, l_{\rm apo})$ space and apply the friends-of-friends (FoF) algorithm to link stars with similar orbits together.

\begin{figure}
	\centering
	\includegraphics[width = \linewidth]{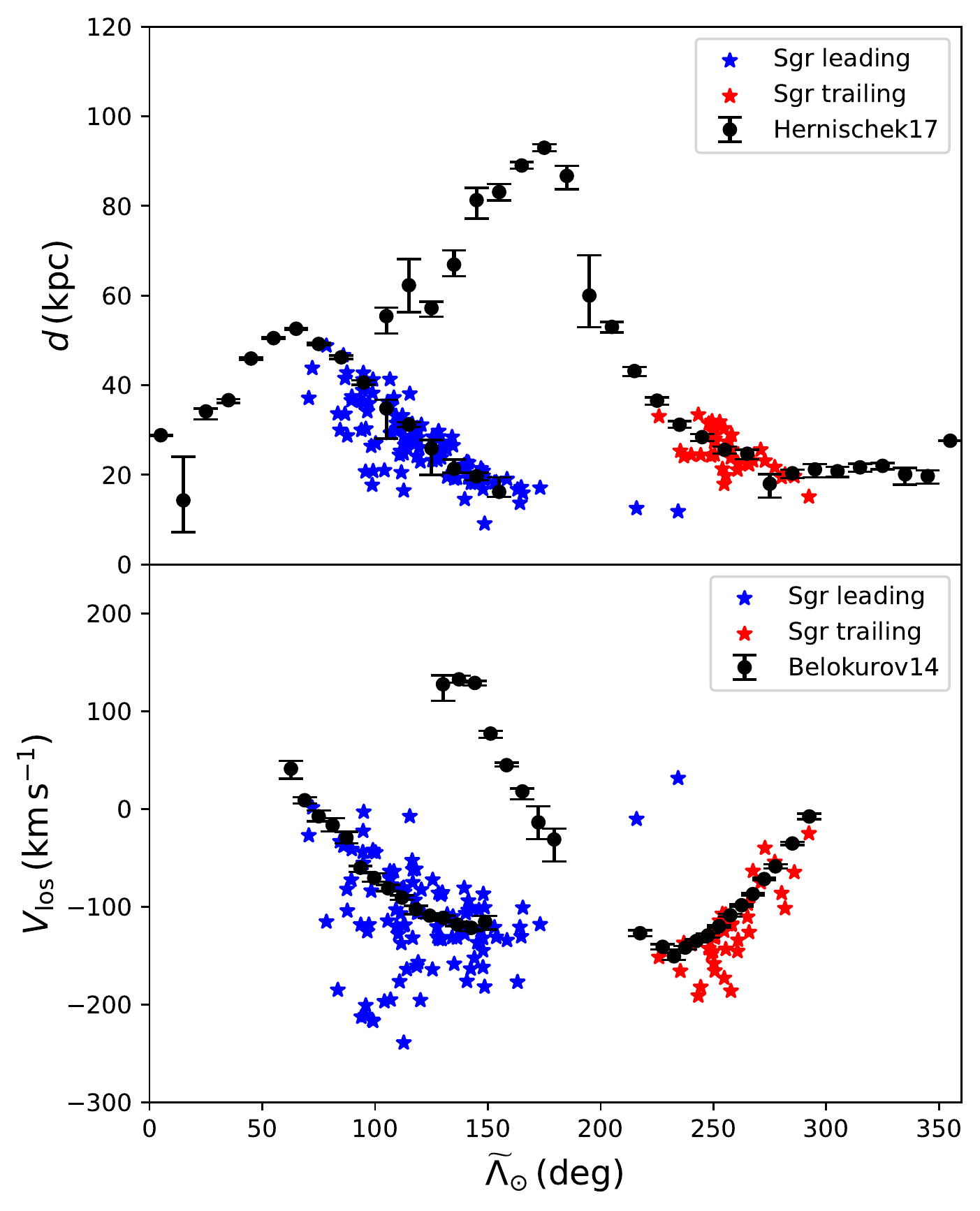}
	\caption{Comparisons with observations of different Sgr data in coordinates of $(\widetilde{\Lambda}_\odot,d)$ and $(\widetilde{\Lambda}_\odot, V_{\rm los})$. The blue and red star symbols represent members of the Sgr leading arm and Sgr trailing arm, respectively. In the top panel, the black dots are from Tables 4 and 5 of Hernitschek et al. (2017) obtained from RR Lyrae stars. In the lower panel, the black dots with error bars are from Tables 3-5 of Belokurov et al. (2014) using Sgr giant stars.}
	\centering
	\label{fig:5}
\end{figure}

\begin{figure}
	\centering
	\includegraphics[width = \linewidth]{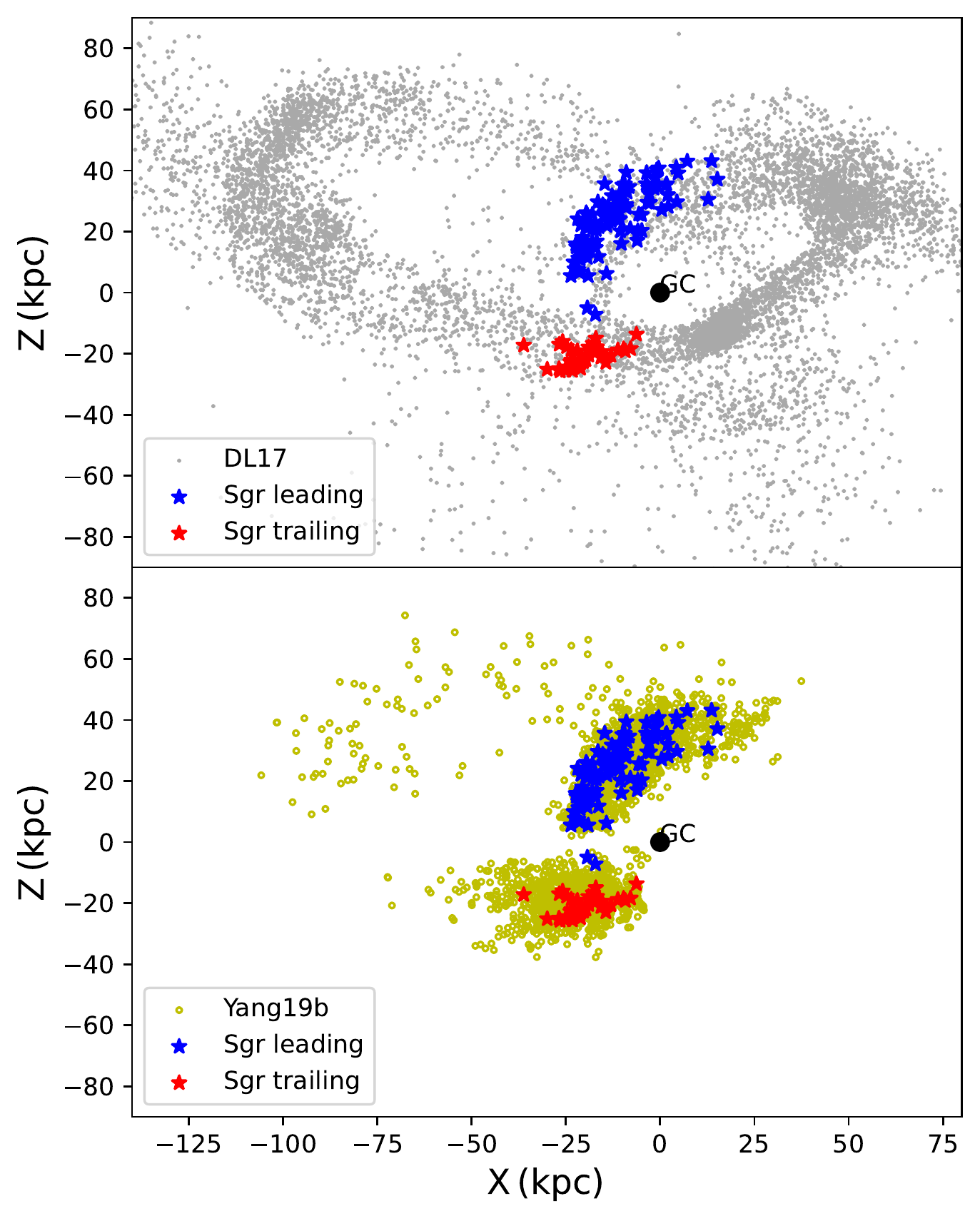}
	\caption{The spatial distributions of Sgr leading arm and trailing arm in the $X$-$Z$ plane. The blue and red star symbols represent members of the Sgr leading arm and Sgr trailing arm, respectively. The gray dots are from the DL17 model and the yellow dots represent the Sgr stream identified in K giants, M giants, and BHBs from Yang et al. (2019b).}
	\label{fig:6}
\end{figure}

\begin{figure*}
	\centering
	\includegraphics[width = \linewidth]{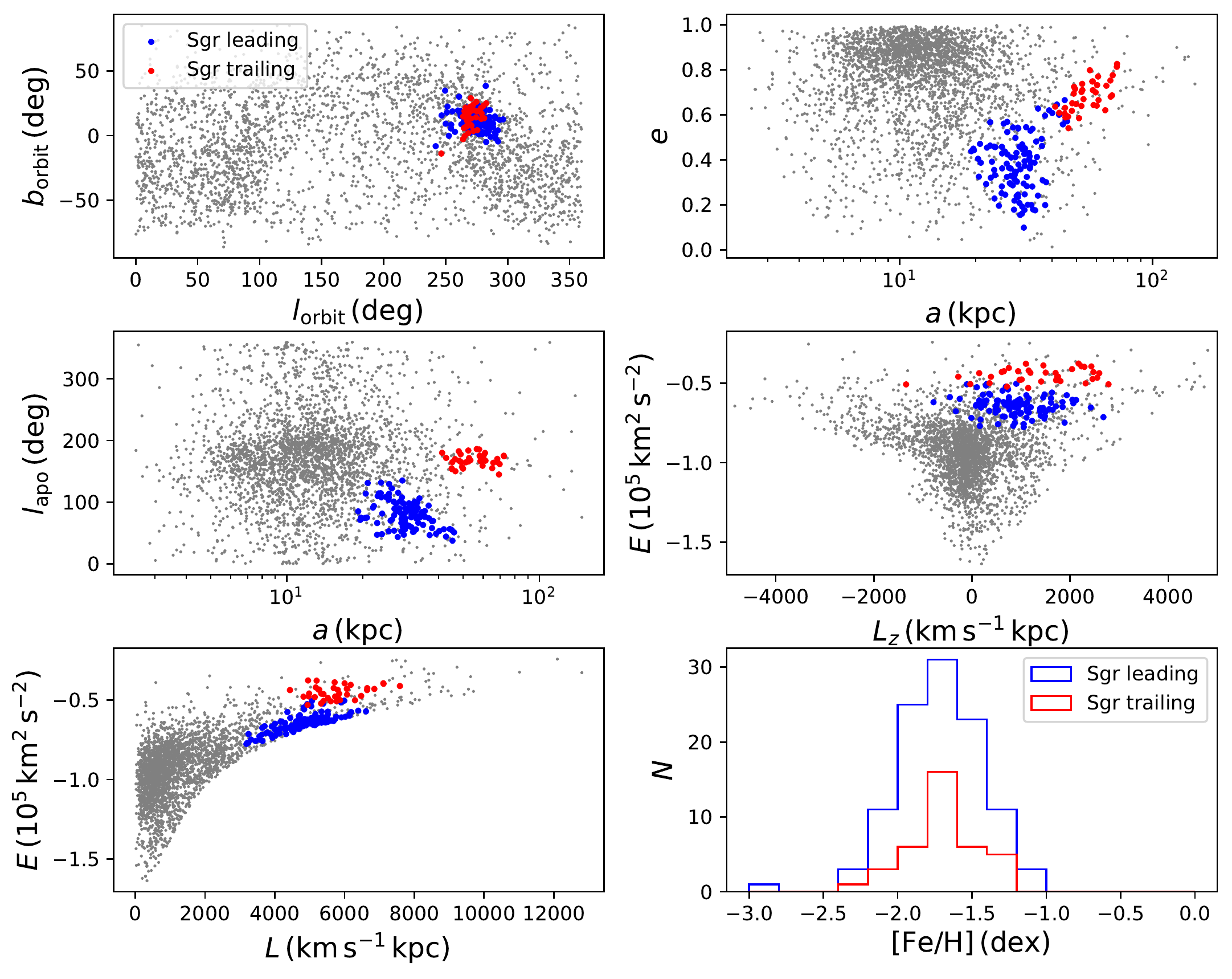}
	\caption{The distributions of our sample and Sgr stream in the IoM space and the metallicity distribution of the Sgr groups. The gray dots represent the total sample. The blue and red dots represent members of the leading arm and the trailing arm, respectively.}
	\label{fig:7}
\end{figure*}

\begin{figure*}
	\centering
	\includegraphics[width = \linewidth]{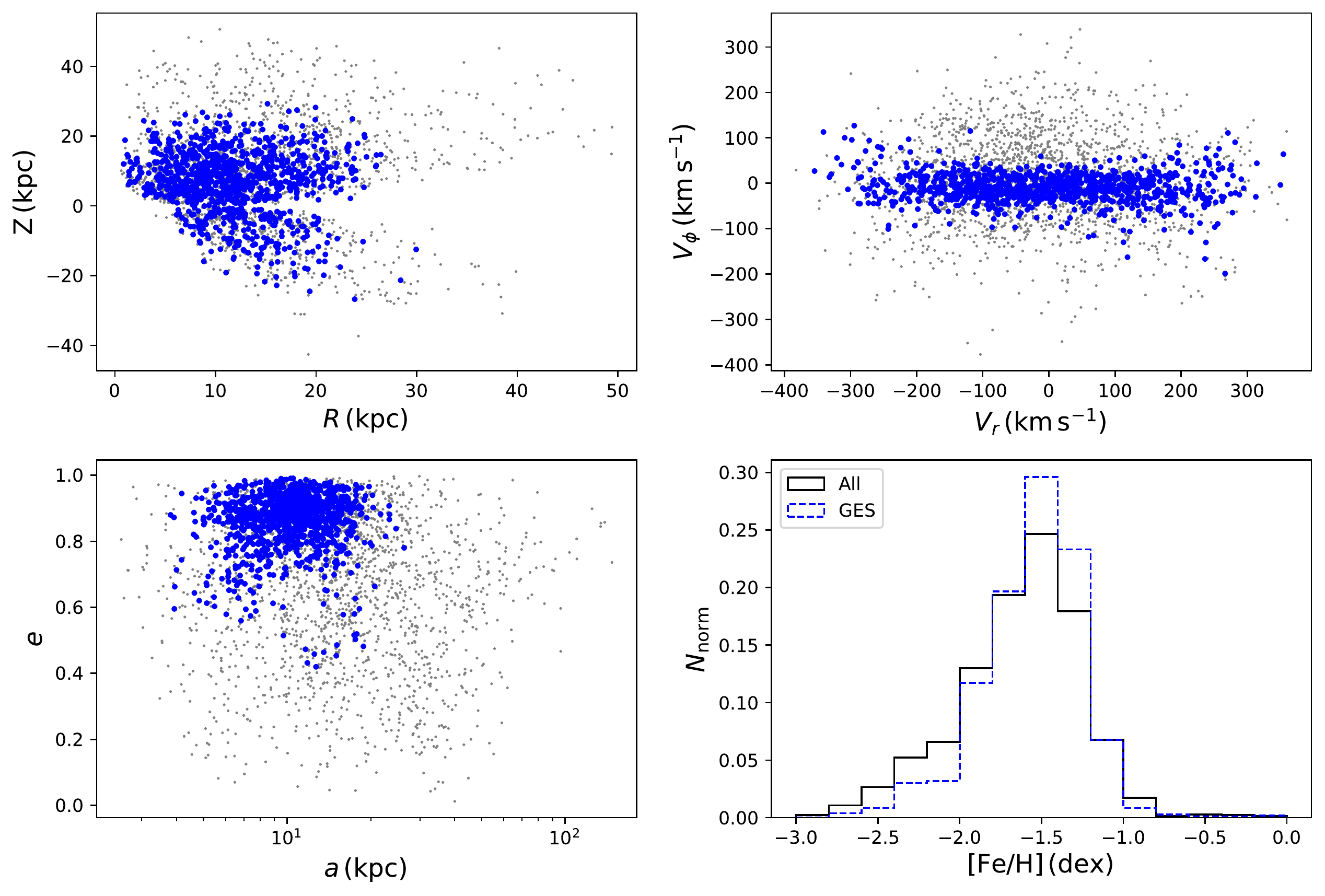}
	\caption{The distributions of our sample and the members of GES in the $(R,z)$, $(V_r, V_\phi)$, $(a,e)$ and the metallicity distribution of the members of GES. The gray dots represent the total sample. The blue dots represent the members of GES.}
	\label{fig:8}
\end{figure*}

\subsection{Integrals of Motion}
It is more effective to identify substructures in the IoM space than in the position-velocity space. In addition, identifying substructures in the IoM space can avoid the influence of the incomplete sky coverage as much as possible. The satellite galaxy will be tidally disrupted and leave stellar debris, like a stellar stream. But under the long-term influence, stars from the satellite galaxy will be scattered in the position-velocity space and are difficult to identify, such as the stars of the Gaia-Enceladus-Sausage (GES; Belokurov et al. 2018; Helmi et al. 2018). In a system with spherical potential without consideration of dynamic friction, there are five IoM parameters: $e$, $a$, $l_{\rm orbit}$, $b_{\rm orbit}$, and $l_{\rm apo}$. Same as Xue et al. (2022, in preparation), we utilise these parameters to characterize the orbit of each star and identify the substructures in these parameter spaces.

The five parameters $(e,a,l_{\rm orbit},b_{\rm orbit}, l_{\rm apo})$ are translated from the total energy $E$ and angular momentum $\boldsymbol{L}$ through combining position-velocity information $(l,b,d,V_{\rm los},V_l,V_b)$ and the Galactic potential. We adopt the Galactic potential composed of a spherical Hernquist bulge (Hernquist 1990), an exponential disk, and a Navarro-Frenk-White (NFW) halo (Navarro et al. 1996).
The direction of the orbital pole $(l_{\rm orbit},b_{\rm orbit})$ is defined as $\tan{(l_{\rm orbit})}=L_y/L_x$ and $\sin{(b_{\rm orbit})}=L_z/L$ by the angular momentum $\boldsymbol{L}$ and its three components $L_x$, $L_y$ and $L_z$ in the Cartesian coordinate system.
The $l_{\rm orbit}$ increases counterclockwise from $0^{\circ}$ to $360^{\circ}$ and the $b_{\rm orbit}$ ranges from $-90^\circ$ to $90^\circ$ (retrograde and prograde orbit respectively). The $b_{\rm orbit}=0$ represents the orbit perpendicular to the Galactic disk. For the $l_{\rm apo}$, the value increases counterclockwise from $0^{\circ}$ to $360^{\circ}$. More precisely, $l_{\rm apo}$ is not an integral of motion because this parameter will change with the orbital period, but it remains constant within one period. So this parameter is also important when we would like to identify the different components of one substructure, e.g. Sgr leading arm and trailing arm.

In order to measure the similarity of two stars in the IoM space, we define the dimensionless separation $\delta_{ij}$ between two stars $i$ and $j$ in these spaces as follows:
\begin{flalign}
\delta_{ij}^2=& \omega_{\theta}\theta_{ij}^2+\omega_{{\Delta}a}(a_i-a_j)^2+\omega_{{\Delta}e}(e_i-e_j)^2 \nonumber \\
&+\omega_{{\Delta}l_{\rm apo}}(l_{{\rm apo},i}-l_{{\rm apo},j})^2,
\end{flalign}
where $\theta_{ij}$ is the circle distance between the direction of the orbital pole $(l_{\rm orbit},b_{\rm orbit})$ of two stars and is calculated by
\begin{flalign}
    \cos{\theta_{ij}}=&\cos{b_{{\rm orbit},i}}\cos{b_{{\rm orbit},j}}\cos{(l_{{\rm orbit},i}-l_{{\rm orbit},j})} \nonumber \\
    &+\sin{b_{{\rm orbit},i}}\sin{b_{{\rm orbit},j}}.
\end{flalign}
The $\omega_{\theta}$, $\omega_{{\Delta}a}$, $\omega_{{\Delta}e}$ and $\omega_{{\Delta}l_{\rm apo}}$ are weights which are used to normalize the corresponding components and are defined as follows:
\begin{gather}
    \omega_\theta=\frac{1}{\langle \theta^2 \rangle}, \omega_{{\Delta}a}=\frac{1}{\langle ({\Delta}a)^2 \rangle},\nonumber \\
    \omega_{{\Delta}e}=\frac{1}{\langle ({\Delta}e)^2 \rangle},  \omega_{{\Delta}l_{\rm apo}}=\frac{1}{\langle ({\Delta}l_{\rm apo})^2 \rangle},
\end{gather}
where $\langle ... \rangle$ refers to the average of all pairs.

\subsection{FoF Algorithm}
We use the FoF group finding algorithm to group stars with similar characteristics. In this algorithm, two stars will be linked together if their separation $\delta$ is within a certain threshold, named as linking length. We calculate the separation $\delta$ between every two of the stars in the $(e,a,l_{\rm orbit},b_{\rm orbit}, l_{\rm apo})$ space. Two stars will be linked as a group if the corresponding $\delta$ is lower than the linking length. Then we link the other stars to the two stars and add those whose separations meet the criteria to the same group. After many iterations, the group is complete until no new stars can be added to the group.

We use a series of different linking lengths to cluster our sample to allow different characteristic sizes of substructures in the algorithm. If we only utilize a small linking length, the FoF algorithm will find a few dense groups which contain few stars and lose some substructures; on the contrary, if we only consider a large linking length, some different characteristic substructures may be linked together due to the lower criteria, which will make the merged substructure unreliable. Therefore, we use a series of different linking lengths to trace the formation and merger history of different groups. Then we set the maximum number of members for a group to remove the unreliable merged groups by checking the results of each linking length. For the same group identified in multiple linking lengths, we only keep the one found in the largest linking length.

The range of linking length we use is from 0.150 to 0.304. The maximum linking length is determined by getting as many reliable members of Sagittarius streams as possible. The maximum physical sizes of each component corresponding to the linking length of 0.150 and 0.304 can be found in Table\,2. The maximum physical size is calculated by the difference component at a given linking length when assuming two stars have the other three identical components in the IoM space.
For example, if two stars have identical values of $l_{\rm orbit}$, $b_{\rm orbit}$, $e$ and $l_{\rm apo}$, a difference of $2.72\,\rm kpc$ in $a$ will cause the $\delta$ of 0.150. For the groups identified by the FoF algorithm, we keep the groups with the number of members not less than 5 and lower than 300 to reduce the influence from the uncertainties of parameters and unreliable merged groups.

\section{Results}
As described in Section 3, we identify substructures with the FoF method in the IoM spaces by combining different linking lengths. Finally, we identify 81 groups (1,411 RRab). The comparison with the known substructures shows that, several groups are associated with the Sagittarius (Sgr) stream (Ibata et al. 2001; Majewski et al. 2003), the Gaia-Enceladus-Sausage (GES; Belokurov et al. 2018; Helmi et al. 2018), the Sequoia (Myeong et al. 2019), and the Helmi streams (Helmi et al. 1999). These substructures have different total angular momentum $L$, $z$-component angular momentum $L_z$, and total energy $E$. The total energy is calculated under the Galactic potential mentioned in Section 3.1. We also compare the $E$ under the Galactic potential we use with that under the potential of McMillan (2017). The mean difference of the energy in different potentials is nearly $0.4\times10^5\,\rm km^2\,s^{-2}$. The properties of these known substructures are shown in Table\,3. As for other groups, we find that three groups are possibly associated with globular clusters NGC 5272, NGC 6656, and NGC 5024, respectively, due to similar kinematic properties. The distributions of these substructures in the $(l,b)$, $(X,Z)$, $(L_z,E)$, and $(L,E)$ space are shown in Fig.\,4, respectively. Besides, there are three remaining groups not linked to any of known substructures.

\begin{figure}
	\centering
	\includegraphics[width = \linewidth]{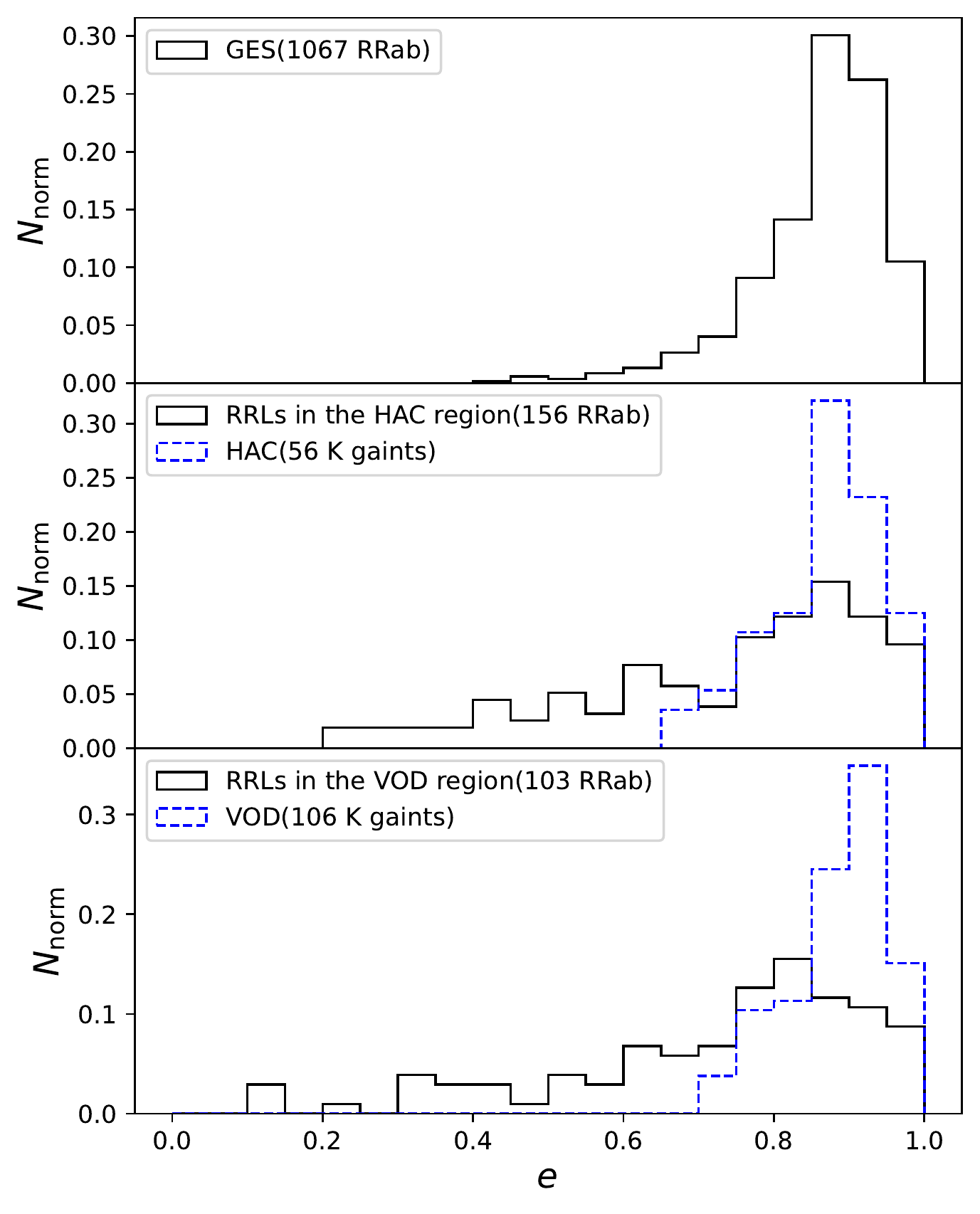}
	\caption{The eccentric distributions of the GES members and the stars in the regions of HAC and VOD. The black histograms represent the GES members and the RRab in the regions of HAC and VOD that don't belong to the GES, respectively. The blue dashed lines represent the K giants belonging to the HAC and VOD from Yang et al. (2019a).}
	\label{fig:9}
\end{figure}

\begin{figure}
	\centering
	\includegraphics[width = \linewidth]{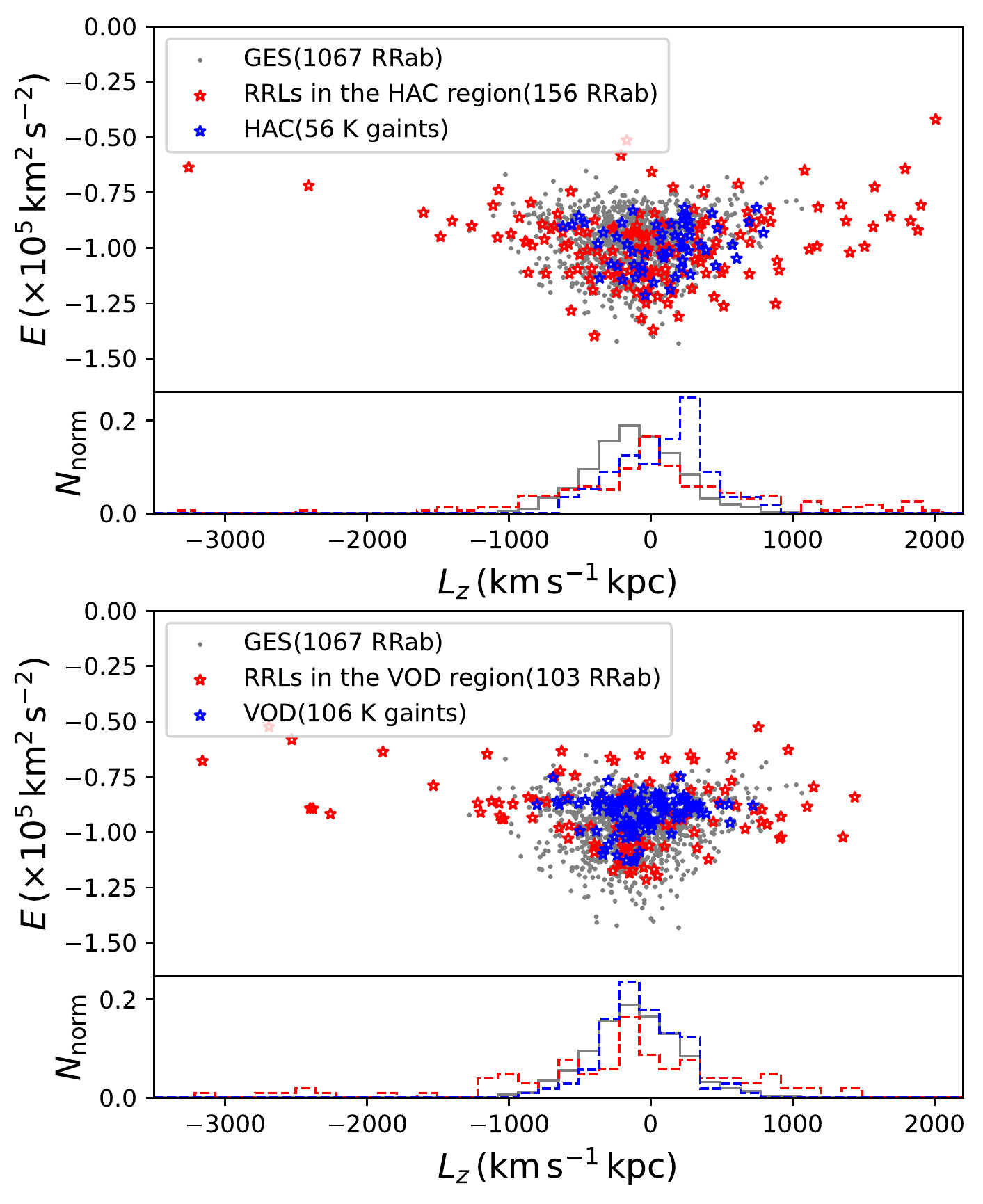}
	\caption{The distributions of the GES members and the stars in the regions of HAC and VOD in the $(L_z,E)$ space. The gray dots represent the RRab belonging to GES. The red stars and dashed lines represent the RRab in the regions of HAC and VOD that don't belong to the GES, respectively. The blue stars and dashed lines represent the K giants belonging to HAC and VOD which are from Yang et al. (2019a).}
	\label{fig:10}
\end{figure}

\subsection{Attributing Groups to Sgr Stream}
Sgr stream is the most prominent substructure in our Galactic halo and a strong tool to study the MW. We compare our groups with the Sgr orbit model (Dierickx \& Loeb 2017; DL17) and RRab Sgr data from PS1 (Hernitscheket al. 2017) and we find three groups (145 RRab) that match well with the DL17 model and observation results from PS1. In these groups, two groups (108 RRab) belong to the Sgr leading arm and another one (37 RRab) belongs to the trailing arm. Fig.\,5 shows the comparisons between these groups with observations of different Sgr data in coordinates of $(\widetilde{\Lambda}_\odot,d)$ and $(\widetilde{\Lambda}_\odot, V_{\rm los})$. $\widetilde{\Lambda}_\odot$ is the longitude in Sgr coordinate system and the definition is the same as that in Belokurov et al. (2014). These groups are associated with other observation results (Belokurov et al. 2014; Hernitscheket al. 2017) and the larger dispersion of radial velocities may be caused by the different Sgr member selection methods. As shown in Fig.\,6, these groups are consistent with the DL17 model and the Sgr streams traced by K giants, M giants, and BHBs (Yang et al. 2019b).

We also study the property of Sgr stream in the IoM space. The distributions in the IoM space and the metallicity distributions of our groups belonging to the Sgr streams are shown in Fig.\,7. We find the directions of the orbital pole are similar in both components. In the $(a,e)$ space, most of the trailing arm members have larger $e$ and $a$ than the leading arm members. But these are a few stars in these two arms with close properties in the $(a,e)$ space, which may cause that these two arms will be linked together if we only consider four IoM parameters $(e,a,l_{\rm orbit},b_{\rm orbit})$.
But these two components will be clearly separated in the $(a,l_{\rm apo})$ space which suggests that the $l_{\rm apo}$ plays an important role when we identify different components in one substructure. In the $(L,E)$ space, the Sgr stream members are distributed in strips with high $L$ and $E$ and the total energy $E$ of trailing arm members is significantly higher than the leading arm. As for the metallicities, the mean and standard deviation are $\rm \langle [Fe/H] \rangle=-1.71\,dex$ and $\sigma_{\rm [Fe/H]}=0.29\,\rm dex$ for RRab belonging to the leading arm, and $\rm \langle [Fe/H] \rangle=-1.69\,dex$ and $\sigma_{\rm [Fe/H]}=0.24\,\rm dex$ for RRab belonging to the trailing arm. The mean metallicity of Sgr RRab is consistent with the results of Yang et al. (2019b) by using BHBs.

\subsection{Attributing Groups to Gaia-Enceladus-Sausage}

\begin{figure}
	\centering
	\includegraphics[width = \linewidth]{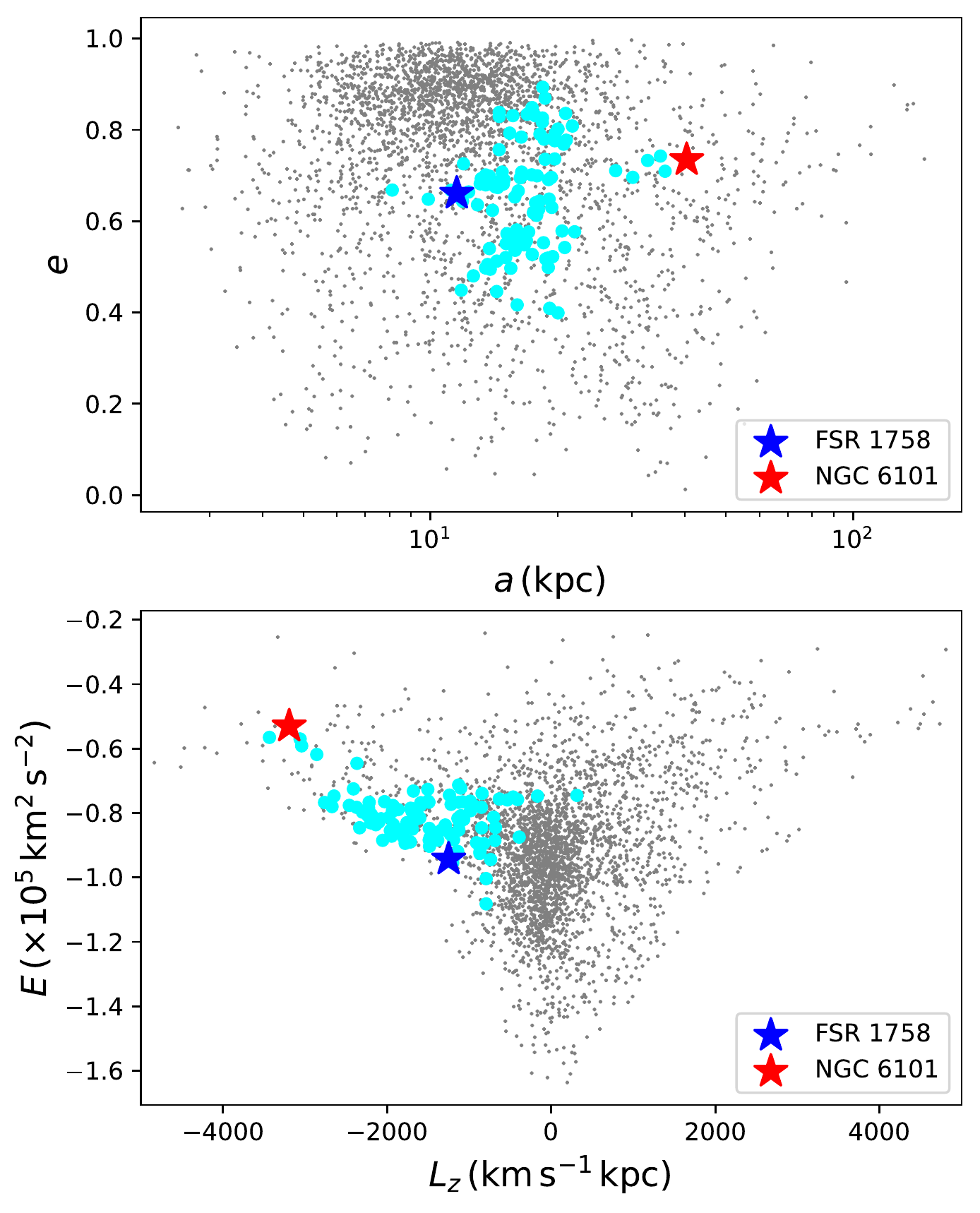}
	\caption{The comparison of the members of Sequoia with GCs NGC 6101 and FSR 1758 in the $(a,e)$ and $(L_z,E)$ space. The gray and cyan dots represent the total sample and the members of Sequoia, respectively. The GCs NGC 6101 and FSR 1758 are shown with red and blue stars.}
	\label{fig:11}
\end{figure}

\begin{figure*}
	\centering
	\includegraphics[width = \linewidth]{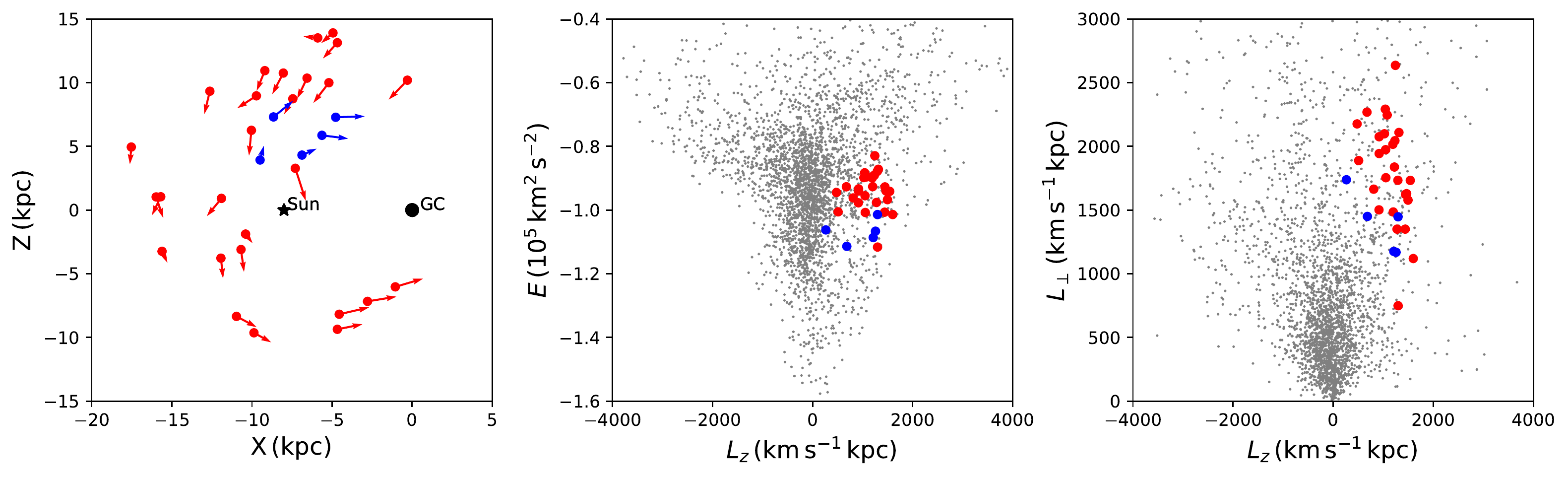}
	\caption{The distributions of the RRab belonging to Helmi streams in the $(X,Z)$, $(L_z,E)$ and $(L_z,L_{\perp})$ space. The gray dots represent our total sample. The red and blue dots represent the RRab members of Helmi streams with counter-clockwise and clockwise motions in the $(X,Z)$ space, respectively. The arrows indicate their moving directions.}
	\label{fig:12}
\end{figure*}

\begin{figure*}
	\centering
	\includegraphics[width = \linewidth]{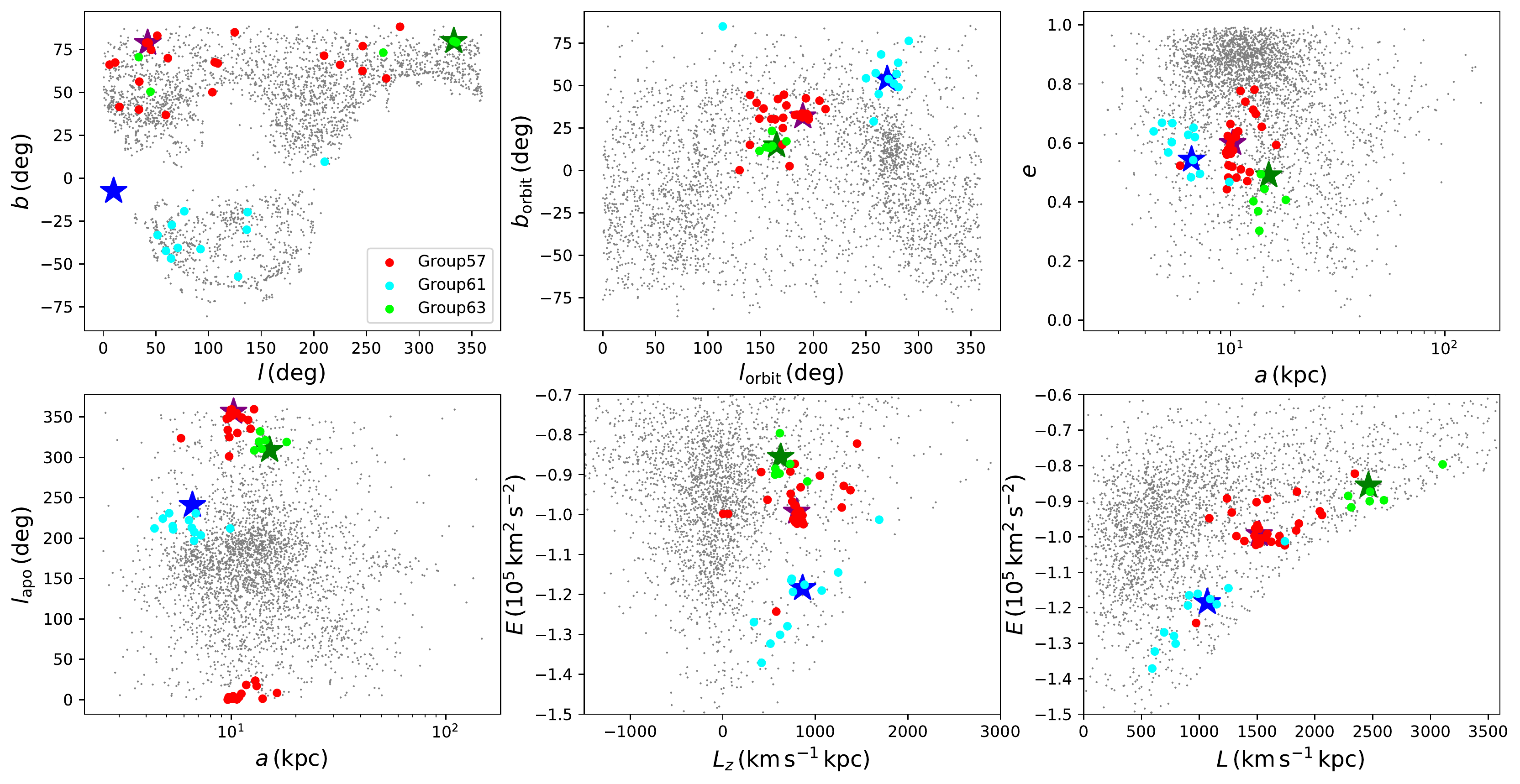}
	\caption{
	The distributions of groups that are likely associated with GCs in $(l,b)$ space and IoM space. The gray dots represent the total sample. The red, cyan, and lime dots represent our groups that are likely consistent with NGC 5272, NGC 6656, and NGC 5024, respectively. The purple, blue and green stars represent the NGC 5272, NGC 6656, and NGC 5024, respectively.}
	\label{fig:13}
\end{figure*}

Belokurov et al. (2018) used a large sample of main-sequence stars within ${\sim}10\,\rm kpc$ from Gaia and SDSS and found that the stellar halo’s velocity ellipsoid was stretched dramatically for stars with $\rm [Fe/H]>-1.7$. They suggested that this property was caused by a major accretion event of a satellite with virial mass $M_{\rm vir}>10^{10}\, M_{\odot}$, the so-called `Gaia Sausage', between 8 and 11 Gyr ago.
Helmi et al. (2018) proposed that the retrograde stars in the halo and some of the low angular momentum stars could be caused by an ancient merger, `Gaia-Enceladus'. The components of the `Gaia-Enceladus' have properties from slight prograde with high eccentric to strongly retrograde. These two events may represent the same merger event due to their properties and we refer to this merger as the Gaia-Enceladus-Sausage (GES).
We select 55 groups (1,062 RRab) associated with GES by utilising the similar criteria as Massari et al. (2019): (i) $-800<L_z<620\,\rm km\,s^{-1}\,kpc$, (ii) $-1.4\times10^5<E<-0.5\times10^5\,\rm km^2\,s^{-2}$, (iii) $L_{\perp}=\sqrt{L_x^2+L_y^2}<3500\,\rm km\,s^{-1}\,kpc$. The criterion of $E$ is adjusted based on the difference of total energy under different Galactic potentials. The group will be selected if not less than half of its members satisfy these criteria. We also select one group (5 RRab) with $e>0.8$ and similar kinematic properties which may be also associated with the GES. The final GES members contain 56 groups (1067 RRab).
The distributions of the members of GES in the $(R,Z)$, $(V_r,V_\phi)$, $(a,e)$ space and the metallicity distributions are shown in Fig.\,8. We find the members of GES are located at a large range in $(R,Z)$ plane and the maximum of $R$ or $Z$ can reach $30\,\rm kpc$. These members have similar properties in the $(V_r,V_\phi)$ plane with `Gaia-Sausage' (Belokurov et al. 2018) which has a very low $V_\phi$ and a large range of $V_r$. 
We estimate the regions and peaks of the semimajor axis $a$ and apocenter distance $r_{\rm apo}$ by utilising the 2.5\%, 50\%, and 97.5\% quantile. Most members of GES span a large range in $a$, from 5 to $18\,\rm kpc$ and the range of $r_{\rm apo}$ is from 10 to $34\,\rm kpc$. The peak values of $a$ and $r_{\rm apo}$ are 11 and $20\,\rm kpc$, respectively. The distributions of $a$ and $r_{\rm,}$ imply that GES is mainly located in the inner halo. Watkins et al. (2009) and Deason et al. (2013) reported that there was a break in the density profile of the Galactic halo at Galactocentric distance ${\sim}20-25\,\rm kpc$. The proportion of the GES in the inner halo is nearly one-third and the peak of the $r_{\rm apo}$ is consistent with the Galactocentric distance of the break, which suggests this break could be caused by the GES.
The mean and standard deviation of metallicities, $\rm \langle [Fe/H] \rangle=-1.56\,dex$ and $\sigma_{\rm [Fe/H]}=0.31\,\rm dex$, are consistent with the metallicity of the GES (Belokurov et al. 2018; Helmi et al. 2018). Also, we find the metallicity distribution of the GES members has no significant difference with all our samples.

The GES accounts for a large part of the groups we have identified and it covers a very large range in the $(l,b)$ space. It's important to study whether this large substructure is related to other known substructures. The Hercules-Aquila Cloud (HAC) is an overdensity that is centered on Galactic longitude $l{\sim}40^{\circ}$ and Galactic latitude $-50^{\circ}<b<50^{\circ}$ (Belokurov et al. 2007). The Virgo Overdensity (VOD) covers over $1000\,\rm deg^2$ and its distance to the Sun is $10\sim20\,\rm kpc$ (Newberg et al. 2007; Bonaca et al. 2012). We find that the majority of RRab belonging to the groups we have identified in these two regions are classified as the GES members, which suggests that the HAC and VOD may have similar kinematic properties to the GES. To verify this possibility, we select stars in the regions of the HAC and VOD defined by Belokurov et al. (2007) and Bonaca et al. (2012), respectively, from all our RRab samples and remove the RRab belonging to the groups we have identified. As shown in Figs.\,9-10, these stars have diffuse distributions in $e$ or $L_z$ space which are mainly contributed by the in-situ stellar halo. Besides, these are also some stars with high $e$ and small $L_z$. These stars identical to the GES members but not be identified could be due to the maximum number of group members we use. However, apart from the stars belonging to the in-situ stellar halo and some similar to the GES member, there is no obvious additional component in these two regions, which implies that these two overdensities have similar kinematic properties to the GES. We also compare the substructures belonging to HAC and VOD from Yang et al. (2019a) which identify the substructures in the position-velocity space by using K giants. As shown in Figs.\,9-10, these two substructures are dominated by stars on highly eccentric orbits and with small $L_z$ which are consistent with the results of Simion et al. (2019). The difference between the GES members and HAC K giants in the $L_z$ distribution may be due to the insufficient HAC K giants. The similar kinematic properties of HAC, VOD, and GES suggest that these substructures may have similar origins.

\subsection{Attributing Groups to Sequoia}
Different from the GES, the Sequoia (Myeong et al. 2019) might be another merger event with a relatively low total mass. The merger debris of the Sequoia has high energy and extra retrograde motion which is clearly separated from the Sausage at nearly zero net angular momentum. We select 12 groups (99 RRab) associated with Sequoia by utilising the similar criteria as Massari et al. (2019): (i) $-3700<L_z<-820\,\rm km\,s^{-1}\,kpc$, (ii) $-1.1\times10^5<E<-0.3\times10^5\,\rm km^2\,s^{-2}$.
We compare these groups with globular clusters (GCs) NGC 6101 and FSR 1758 which are associated with the Sequoia (Myeong et al. 2019) in IoM space in Fig.\,11. The kinematic properties of FSR 1758 are from Simpson (2019). We find one group with $E>-0.7{\times}10^5\,\rm km^2\,s^{-2}$ has a larger $e$ and $a$, with average values of 0.72 and $32\,\rm kpc$ which may be associated with NGC 6101. Others with lower $E$ are likely associated with FSR 1758. There are a few stars with very low $L_z$ which may be contaminations due to the large linking length.
The mean metallicity of all members $\rm \langle [Fe/H] \rangle=-1.70\,dex$ is lower than the GES stars, which is consistent with the more metal-poor Sequoia (Myeong et al. 2019).

\subsection{Attributing Groups to Helmi Streams}
Helmi streams were the debris streams that were identified in the solar neighborhood (Helmi et al. 1999). Using Gaia DR2, Koppelman et al. (2019) found nearly 600 new members of the Helmi streams up to a distance of 5 kpc from the Sun. They found the peak of the metallicities was near $-1.5\,\rm dex$ and the age range was from $\approx$11 to $13\,\rm Gyr$, which confirmed that the Helmi streams originated from a dwarf galaxy. Combining the criteria of Koppelman et al. (2019) and Massari et al. (2019), we find 4 groups (32 RRab) associated with the Helmi streams by utilising the following criteria: (i) $620<L_z<1700\,\rm km\,s^{-1}\,kpc$, (ii) $1000<L_{\perp}<3200\,\rm km\,s^{-1}\,kpc$, (iii) $E<-0.6\times10^5<E<-0.3\times10^5\,\rm km^2\,s^{-2}$. The spatial distributions of these RRab are shown in Fig.\,12. These RRab cover a large heliocentric distance range of $3-15\,\rm kpc$. The mean orbit radii at apocentre and pericentre are 16 and $7\,\rm kpc$ which are in agreement with the results of Helmi et al. (1999) estimated by the red giants and RRLs within $1\,\rm kpc$ of the Sun. The maximum of the orbital radii at apocentre is $23\,\rm kpc$, which means that most of Helmi streams within a Galactocentric distance of $23\,\rm kpc$. Same as the result of Koppelman et al. (2019), the members of Helmi streams can be divided into two components: three groups (27 RRab) with counter-clockwise motions in the $(X,Z)$ space and one group (5 RRab) with clockwise motions in the $(X,Z)$ space. For all members of Helmi streams, the mean and standard deviation of metallicities, $\rm \langle [Fe/H] \rangle=-1.54\,dex$ and $\sigma_{\rm [Fe/H]}=0.45\,\rm dex$, are consistent with the metallicity properties of the Helmi streams (Koppelman et al. 2019).

\subsection{Attributing Groups to Globular Clusters}
Yuan et al. (2020) found several dynamically tagged groups were associated with known MW globular clusters and a new stream that was associated with a pair of GCs (NGC 5024 and NGC 5053), by using BHBs and RRLs. We also compare the remaining groups with GCs in the IoM space and metallicities. The GCs are from Vasiliev (2019) which determines the mean proper motions for 150 GCs using Gaia DR2. We find that three groups, Group57 (31 RRab), Group61 (12 RRab), and Group63 (6 RRab), are likely associated with the GCs, NGC 5272, NGC 6656, and NGC 5024, respectively, due to similar kinematic properties. The name of the group represents the group's ID. The distributions of these groups in the spatial space and IoM space are shown in Fig.\,13. The mean and standard deviation of metallicities of Group57 are $-1.76\,\rm dex$ and $0.34\,\rm dex$ which are consistent with the metallicities of NGC 5272 $-1.50\,\rm dex$ (Harris 1996, edition 2010). The mean metallicity of Group61 and Group63 are $-1.43$ and $-1.78\,\rm dex$, respectively, which are larger than the metallicities of NGC 6656 and NGC 5024, $-1.70$ and $-2.10\,\rm dex$ (Harris 1996, edition 2010).

\begin{figure*}
	\centering
	\includegraphics[width = \linewidth]{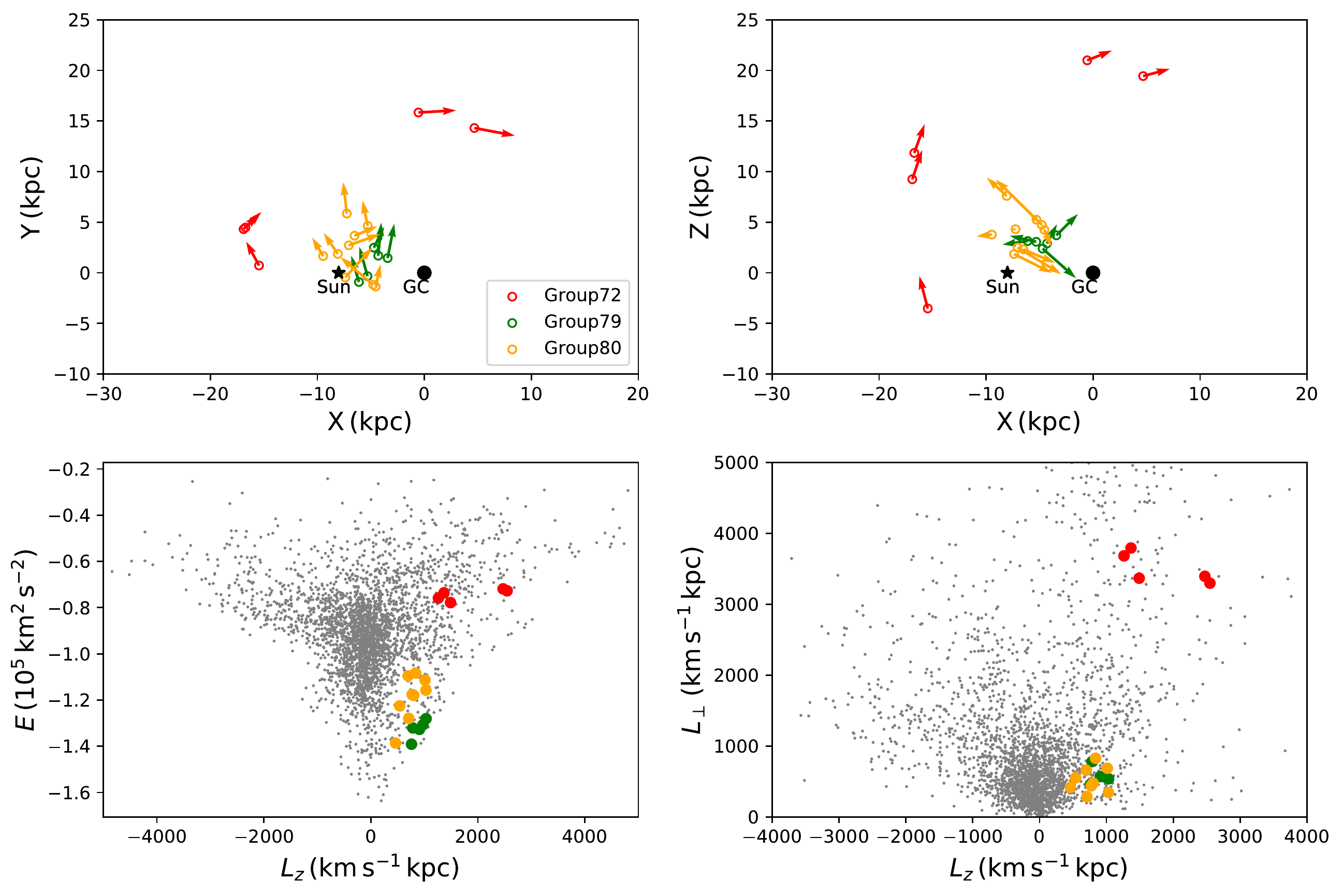}
	\caption{
	The distributions of the remaining groups in the $(\rm X,Y)$, $(\rm X,Z)$, $(L_z,E)$ and $(L_z,L_{\perp})$ space. The gray dots represent the total sample and the red, green, and orange dots represent the three remaining groups. The arrows in the left panel show their moving directions and velocity amplitudes.}
	\label{fig:14}
\end{figure*}

\subsection{The remaining groups}
There are three remaining groups that are likely not related to the known large substructures or GCs. Fig.\,14 illustrates the distributions of these groups in the $(X,Y)$, $(X,Z)$, $(L_z,E)$, and $(L_x,L_y)$ space. In these groups, Group72 has very special kinematic properties.
The distribution of this group in the $(L_z,E)$ space is similar to the Sgr leading arm but the $L_y$ is positive which is very different from Sgr leading arm. We also compare this group with the Helmi streams (Helmi et al. 1999; Koppelman et al. 2019). The mean $L_{\perp}$ of this group is $3,508\,\rm km\,s^{-1}\,kpc$ which is higher than the Helmi streams. The mean and standard deviation of metallicities are $-2.13\,\rm dex$ and $0.24\,\rm dex$, respectively. So we suggest that this group may be a new substructure. 
The orbital parameters of all substructures we identified is shown in Table\,4.

\begin{table*}
    
	\centering
	\caption{The orbital parameters of all the substructures we have identified.}
	\label{tab:4}
	\begin{threeparttable}
	%\resizebox{2\columnwidth}{!}{
	\begin{tabular}{cccccccccccc} % four columns, alignment for each
		\hline
index$\rm ^a$ & $l_{\rm orbit}$ & $b_{\rm orbit}$ & $a$ & $e$ & $l_{\rm apo}$ & $E$ & $L$ & $L_x$ & $L_y$ & $L_z$ & label$\rm ^b$\\
 & $\rm (deg)$ & $\rm (deg)$ & $\rm (kpc)$ &  &  $\rm (deg)$ & $\rm (km^2\,s^{-2})$ & $\rm (km\,s^{-1}\,kpc)$ & $\rm (km\,s^{-1}\,kpc)$ & $\rm (km\,s^{-1}\,kpc)$ & $\rm (km\,s^{-1}\,kpc)$ & \\
\hline
1240 & 290.22 & 0.85 & 29.74 & 0.21 & 84.68 & $-64189.78$ & 5338.39 & 1845.04 & $-5008.79$ & 79.15 & 0 \\
1238 & 255.37 & 18.61 & 26.14 & 0.24 & 72.03 & $-68174.50$ & 4689.75 & $-1122.70$ & $-4300.46$ & 1496.47 & 0 \\
1626 & 278.16 & 8.86 & 30.91 & 0.29 & 86.83 & $-62818.50$ & 5330.85 & 747.58 & $-5213.92$ & 821.02 & 0 \\
1630 & 270.29 & 11.15 & 30.52 & 0.47 & 61.67 & $-62637.13$ & 4666.61 & 23.38 & $-4578.54$ & 902.03 & 0 \\
1430 & 283.55 & 6.85 & 31.83 & 0.32 & 81.85 & $-61849.07$ & 5388.02 & 1253.21 & $-5200.67$ & 642.85 & 0 \\
1193 & 276.09 & 23.40 & 23.58 & 0.46 & 114.49 & $-70761.57$ & 3747.68 & 365.13 & $-3420.07$ & 1488.25 & 0 \\
1635 & 254.30 & 5.50 & 29.07 & 0.30 & 47.14 & $-64703.93$ & 5020.91 & $-1352.33$ & $-4811.34$ & 481.38 & 0 \\
219 & 276.68 & 5.33 & 29.35 & 0.24 & 71.79 & $-64536.36$ & 5202.41 & 602.91 & $-5144.69$ & 483.49 & 0 \\
1184 & 249.58 & 34.66 & 20.78 & 0.47 & 75.07 & $-74804.97$ & 3319.64 & $-952.66$ & $-2558.94$ & 1887.93 & 0 \\
222 & 271.10 & 20.23 & 27.25 & 0.30 & 85.27 & $-66740.95$ & 4756.24 & 86.00 & $-4461.99$ & 1644.70 & 0 \\
		\hline
	\end{tabular}
	\begin{tablenotes}
	\footnotesize
	\item[] \textbf{Notes}
	\item[a] Index same as that in Table\,1.
	\item[b] 0-Sgr leading arm; 1-Sgr trailing arm; 2-GES; 3-Sequoia; 4-Helmi streams;
		5-likely NGC 5272; 6-likely NGC 6656; 7-likely NGC 5024; 
		8,9,10-unknown groups: Group72, Group79, Group80.
	\item[] (This table is available in its entirety in machine-readable form.)
	\end{tablenotes}
	\end{threeparttable}
\end{table*}

\section{Summary}
In this work, we identify substructures in the MW using a relatively large RRab catalogue with 6D position-velocity information and metallicities. By combining the recently published RRL catalogue from photometric surveys with spectroscopic data from LAMOST and SDSS datasets, we enlarge the RRab dataset of Liu et al. (2020) with radial velocity, metallicity, and distance, and obtain the proper motions from Gaia EDR3.
By comparing the kinematic properties of RRab with different metallicities, we find our RRab sample can be roughly divided into the metal-rich (disk) and the metal-poor (halo) components at $\rm [Fe/H]{\sim}-0.8\,dex$. After removing most of the disk RRab, we utilize FoF algorithm to identify substructures with similar characteristics in the IoM space $(e,a,l_{\rm orbit},b_{\rm orbit}, l_{\rm apo})$. 
In total, we identify 81 groups (1,411 RRab). By comparing with the known substructures, we find several groups associated with known substructures, Sgr Stream (145 RRab), Gaia-Enceladus-Sausage (1,067 RRab), the Sequoia (99 RRab), and the Helmi streams (32 RRab). Our Sgr groups suggest that the projection of $X$-axis on the orbital plane $l_{\rm apo}$ is an effective parameter to distinguish the Sgr leading arm from the trailing arm. For GES, the proportion of this accretion event is very large in our groups and GES members are located within a large area in the $(R,Z)$ plane. In addition, the range of $r_{\rm apo}$ for GES members is from 10 to $34\,\rm kpc$, which suggests that the GES is mainly distributed in the inner halo. The near one-third proportion of the GES and the peak value $20\,\rm kpc$ of the $r_{\rm apo}$ suggest that GES could account for the break in the density profile of the Galactic halo at the Galactocentric distance of ${\sim}20-25\,\rm kpc$.
It is also significant to study the relations between GES and other known substructures in the inner halo. From the comparison with known substructures in the inner halo, the Hercules-Aquila Cloud and Virgo Overdensity, the similar kinematic properties suggest that the Gaia-Enceladus-Sausage and these two substructures may have similar origins.
Besides these large substructures, we compare the remaining groups with the GCs and find that the kinematic properties of 31, 12, and 6 RRab stars are similar to NGC 5272, NGC 6656, and NGC 5024 respectively, which may imply the association between them. Finally, we also find three remaining unknown substructures, one of which has large angular momentum and a mean metallicity of $\rm -2.13\,dex$ and may be a possible new substructure. This unknown substructure may need more data to confirm.

\section*{Acknowledgements}

We thank Dr. Shuo Zhang for her kind help.
This work was funded by the National Key R\&D Program of China No. 2019YFA0405500, the science research grants from the China Manned Space Project with No. CMS-CSST-2021-B05, and the National Natural Science Foundation of China (NSFC) under grant No.11973001, 12090040, and 12090044.
X.X.X is supported by NSFC under grant No. 11988101, 11873052, 11890694, and the science research grants from the China Manned Space Project with No. CMS-CSST-2021-B03.
Y.H. is supported by NSFC under grant No. 11903027, 11833006, 11811530289, and U1731108, and the Yunnan University grant C176220100006.
G.C.L is supported by NSFC under grant No. U1731108.

This work has made use of data products from the Guo Shou Jing Telescope (the Large Sky Area Multi-Object Fibre Spectroscopic Telescope, LAMOST). LAMOST is a National Major Scientific
Project built by the Chinese Academy of Sciences. Funding for the project has been provided by the National Development and Reform Commission. LAMOST is operated and managed by the National Astronomical Observatories, Chinese Academy of Sciences.

Funding for the Sloan Digital Sky Survey IV has been provided by the Alfred P. Sloan Foundation, the U.S. Department of Energy Office of Science, and the Participating Institutions. SDSS acknowledges support and resources from the Center for High-Performance Computing at the University of Utah. The SDSS web site is www.sdss.org.

SDSS is managed by the Astrophysical Research Consortium for the Participating Institutions of the SDSS Collaboration including the Brazilian Participation Group, the Carnegie Institution for Science, Carnegie Mellon University, Center for Astrophysics | Harvard \& Smithsonian (CfA), the Chilean Participation Group, the French Participation Group, Instituto de Astrofísica de Canarias, The Johns Hopkins University, Kavli Institute for the Physics and Mathematics of the Universe (IPMU) / University of Tokyo, the Korean Participation Group, Lawrence Berkeley National Laboratory, Leibniz Institut für Astrophysik Potsdam (AIP), Max-Planck-Institut für Astronomie (MPIA Heidelberg), Max-Planck-Institut für Astrophysik (MPA Garching), Max-Planck-Institut für Extraterrestrische Physik (MPE), National Astronomical Observatories of China, New Mexico State University, New York University, University of Notre Dame, Observatório Nacional / MCTI, The Ohio State University, Pennsylvania State University, Shanghai Astronomical Observatory, United Kingdom Participation Group, Universidad Nacional Autónoma de México, University of Arizona, University of Colorado Boulder, University of Oxford, University of Portsmouth, University of Utah, University of Virginia, University of Washington, University of Wisconsin, Vanderbilt University, and Yale University.

This work has made use of data from the European Space Agency (ESA) mission
{\it Gaia} (\url{https://www.cosmos.esa.int/gaia}), processed by the {\it Gaia}
Data Processing and Analysis Consortium (DPAC,
\url{https://www.cosmos.esa.int/web/gaia/dpac/consortium}). Funding for the DPAC
has been provided by national institutions, in particular the institutions
participating in the {\it Gaia} Multilateral Agreement.

\section*{DATA AVAILABILITY}
The data underlying this article are available in the article and in its online supplementary material.





\begin{thebibliography}{}
\bibitem[Alam et al.(2015)]{2015ApJS..219...12A} Alam, S., Albareti, F.~D., Allende Prieto, C., et al.\ 2015, \apjs, 219, 12. 
\bibitem[Belokurov et al.(2007)]{2007ApJ...657L..89B} Belokurov, V., Evans, N.~W., Bell, E.~F., et al.\ 2007, \apjl, 657, L89.
\bibitem[Belokurov et al.(2014)]{2014MNRAS.437..116B} Belokurov, V., Koposov, S.~E., Evans, N.~W., et al.\ 2014, \mnras, 437, 116. 
\bibitem[Belokurov et al.(2018)]{2018MNRAS.478..611B} Belokurov, V., Erkal, D., Evans, N.~W., et al.\ 2018, \mnras, 478, 611. 
\bibitem[Blumenthal et al.(1984)]{1984Natur.311..517B} Blumenthal, G.~R., Faber, S.~M., Primack, J.~R., et al.\ 1984, \nat, 311, 517. 
\bibitem[Bonaca et al.(2012)]{2012AJ....143..105B} Bonaca, A., Juri{\'c}, M., Ivezi{\'c}, {\v{Z}}., et al.\ 2012, \aj, 143, 105.
\bibitem[Bovy(2015)]{2015ApJS..216...29B} Bovy, J.\ 2015, \apjs, 216, 29. 
\bibitem[Bullock et al.(2001)]{2001ApJ...548...33B} Bullock, J.~S., Kravtsov, A.~V., \& Weinberg, D.~H.\ 2001, \apj, 548, 33. 
\bibitem[Bullock \& Johnston(2005)]{2005ApJ...635..931B} Bullock, J.~S. \& Johnston, K.~V.\ 2005, \apj, 635, 931. 
\bibitem[Chadid et al.(2017)]{2017ApJ...835..187C} Chadid, M., Sneden, C., \& Preston, G.~W.\ 2017, \apj, 835, 187. 
\bibitem[Chambers et al.(2016)]{2016arXiv161205560C} Chambers, K.~C., Magnier, E.~A., Metcalfe, N., et al.\ 2016, arXiv:1612.05560
\bibitem[Clementini et al.(2019)]{2019A&A...622A..60C} Clementini, G., Ripepi, V., Molinaro, R., et al.\ 2019, \aap, 622, A60. 
\bibitem[Cooper et al.(2010)]{2010MNRAS.406..744C} Cooper, A.~P., Cole, S., Frenk, C.~S., et al.\ 2010, \mnras, 406, 744. 
\bibitem[Cui et al.(2012)]{2012RAA....12.1197C} Cui, X.-Q., Zhao, Y.-H., Chu, Y.-Q., et al.\ 2012, Research in Astronomy and Astrophysics, 12, 1197. 
\bibitem[Deason et al.(2013)]{2013ApJ...763..113D} Deason, A.~J., Belokurov, V., Evans, N.~W., et al.\ 2013, \apj, 763, 113. 
\bibitem[Deng et al.(2012)]{2012RAA....12..735D} Deng, L.-C., Newberg, H.~J., Liu, C., et al.\ 2012, Research in Astronomy and Astrophysics, 12, 735. 
\bibitem[Dierickx \& Loeb(2017)]{2017ApJ...836...92D} Dierickx, M.~I.~P. \& Loeb, A.\ 2017, \apj, 836, 92. 
\bibitem[Drake et al.(2013a)]{2013ApJ...763...32D} Drake, A.~J., Catelan, M., Djorgovski, S.~G., et al.\ 2013, \apj, 763, 32. 
\bibitem[Drake et al.(2013b)]{2013ApJ...765..154D} Drake, A.~J., Catelan, M., Djorgovski, S.~G., et al.\ 2013, \apj, 765, 154. 
\bibitem[Drake et al.(2014)]{2014ApJS..213....9D} Drake, A.~J., Graham, M.~J., Djorgovski, S.~G., et al.\ 2014, \apjs, 213, 9. 
\bibitem[Drake et al.(2017)]{2017MNRAS.469.3688D} Drake, A.~J., Djorgovski, S.~G., Catelan, M., et al.\ 2017, \mnras, 469, 3688. 
\bibitem[Fabricius et al.(2021)]{2021A&A...649A...5F} Fabricius, C., Luri, X., Arenou, F., et al.\ 2021, \aap, 649, A5. 
\bibitem[Freeman \& Bland-Hawthorn(2002)]{2002ARA&A..40..487F} Freeman, K. \& Bland-Hawthorn, J.\ 2002, \araa, 40, 487. 
\bibitem[Gaia Collaboration et al.(2016)]{2016A&A...595A...1G} Gaia Collaboration, Prusti, T., de Bruijne, J.~H.~J., et al.\ 2016, \aap, 595, A1.
\bibitem[Gaia Collaboration et al.(2018)]{2018A&A...616A...1G} Gaia Collaboration, Brown, A.~G.~A., Vallenari, A., et al.\ 2018, \aap, 616, A1. 
\bibitem[Gaia Collaboration et al.(2021)]{2021A&A...649A...1G} Gaia Collaboration, Brown, A.~G.~A., Vallenari, A., et al.\ 2021, \aap, 649, A1. 
\bibitem[Harris(1996)]{1996AJ....112.1487H} Harris, W.~E.\ 1996, \aj, 112, 1487.
\bibitem[Haywood et al.(2018)]{2018ApJ...863..113H} Haywood, M., Di Matteo, P., Lehnert, M.~D., et al.\ 2018, \apj, 863, 113. 
\bibitem[Helmi et al.(1999)]{1999Natur.402...53H} Helmi, A., White, S.~D.~M., de Zeeuw, P.~T., et al.\ 1999, \nat, 402, 53. 
\bibitem[Helmi et al.(2018)]{2018Natur.563...85H} Helmi, A., Babusiaux, C., Koppelman, H.~H., et al.\ 2018, \nat, 563, 85. 
\bibitem[Helmi(2020)]{2020ARA&A..58..205H} Helmi, A.\ 2020, \araa, 58, 205.
\bibitem[Hernitschek et al.(2017)]{2017ApJ...850...96H} Hernitschek, N., Sesar, B., Rix, H.-W., et al.\ 2017, \apj, 850, 96. 
\bibitem[Hernquist(1990)]{1990ApJ...356..359H} Hernquist, L.\ 1990, \apj, 356, 359. 
\bibitem[Ibata et al.(1994)]{1994Natur.370..194I} Ibata, R.~A., Gilmore, G., \& Irwin, M.~J.\ 1994, \nat, 370, 194. 
\bibitem[Ibata et al.(1995)]{1995MNRAS.277..781I} Ibata, R.~A., Gilmore, G., \& Irwin, M.~J.\ 1995, \mnras, 277, 781. 
\bibitem[Ibata et al.(2001)]{2001ApJ...551..294I} Ibata, R., Lewis, G.~F., Irwin, M., et al.\ 2001, \apj, 551, 294. 
\bibitem[Jayasinghe et al.(2019)]{2019MNRAS.485..961J} Jayasinghe, T., Stanek, K.~Z., Kochanek, C.~S., et al.\ 2019, \mnras, 485, 961.
\bibitem[Keller et al.(2008)]{2008ApJ...678..851K} Keller, S.~C., Murphy, S., Prior, S., et al.\ 2008, \apj, 678, 851. 
\bibitem[Kerr \& Lynden-Bell(1986)]{1986MNRAS.221.1023K} Kerr, F.~J. \& Lynden-Bell, D.\ 1986, \mnras, 221, 1023. 
\bibitem[Klement et al.(2008)]{2008ApJ...685..261K} Klement, R., Fuchs, B., \& Rix, H.-W.\ 2008, \apj, 685, 261. 
\bibitem[Klement et al.(2009)]{2009ApJ...698..865K} Klement, R., Rix, H.-W., Flynn, C., et al.\ 2009, \apj, 698, 865. 
\bibitem[Koppelman et al.(2019)]{2019A&A...625A...5K} Koppelman, H.~H., Helmi, A., Massari, D., et al.\ 2019, \aap, 625, A5. 
\bibitem[Layden et al.(1996)]{1996AJ....112.2110L} Layden, A.~C., Hanson, R.~B., Hawley, S.~L., et al.\ 1996, \aj, 112, 2110.
\bibitem[Liu et al.(2014)]{2014IAUS..298..310L} Liu, X.-W., Yuan, H.-B., Huo, Z.-Y., et al.\ 2014, Setting the scene for Gaia and LAMOST, 298, 310. 
\bibitem[Liu et al.(2020)]{2020ApJS..247...68L} Liu, G.-C., Huang, Y., Zhang, H.-W., et al.\ 2020, \apjs, 247, 68. 
\bibitem[Luo et al.(2021)]{2021ASPC..529..147L} Luo, C., Liu, C., Zhang, X., et al.\ 2021, RR Lyrae/Cepheid 2019: Frontiers of Classical Pulsators, 529, 147
\bibitem[Majewski et al.(2003)]{2003ApJ...599.1082M} Majewski, S.~R., Skrutskie, M.~F., Weinberg, M.~D., et al.\ 2003, \apj, 599, 1082. 
\bibitem[Marsakov et al.(2018)]{2018ARep...62...50M} Marsakov, V.~A., Gozha, M.~L., \& Koval, V.~V.\ 2018, Astronomy Reports, 62, 50.
\bibitem[Massari et al.(2019)]{2019A&A...630L...4M} Massari, D., Koppelman, H.~H., \& Helmi, A.\ 2019, \aap, 630, L4.
\bibitem[McMillan(2017)]{2017MNRAS.465...76M} McMillan, P.~J.\ 2017, \mnras, 465, 76.
\bibitem[Morrison et al.(2009)]{2009ApJ...694..130M} Morrison, H.~L., Helmi, A., Sun, J., et al.\ 2009, \apj, 694, 130. 
\bibitem[Myeong et al.(2018)]{2018ApJ...856L..26M} Myeong, G.~C., Evans, N.~W., Belokurov, V., et al.\ 2018, \apjl, 856, L26. 
\bibitem[Myeong et al.(2019)]{2019MNRAS.488.1235M} Myeong, G.~C., Vasiliev, E., Iorio, G., et al.\ 2019, \mnras, 488, 1235. 
\bibitem[Navarro et al.(1996)]{1996ApJ...462..563N} Navarro, J.~F., Frenk, C.~S., \& White, S.~D.~M.\ 1996, \apj, 462, 563. 
\bibitem[Newberg et al.(2002)]{2002ApJ...569..245N} Newberg, H.~J., Yanny, B., Rockosi, C., et al.\ 2002, \apj, 569, 245. 
\bibitem[Newberg et al.(2007)]{2007ApJ...668..221N} Newberg, H.~J., Yanny, B., Cole, N., et al.\ 2007, \apj, 668, 221. 
\bibitem[Peebles(1974)]{1974ApJ...189L..51P} Peebles, P.~J.~E.\ 1974, \apjl, 189, L51.
\bibitem[Prudil et al.(2020)]{2020MNRAS.492.3408P} Prudil, Z., D{\'e}k{\'a}ny, I., Grebel, E.~K., et al.\ 2020, \mnras, 492, 3408. 
\bibitem[Reid(1993)]{1993ARA&A..31..345R} Reid, M.~J.\ 1993, \araa, 31, 345. 
\bibitem[Samus' et al.(2017)]{2017ARep...61...80S} Samus', N.~N., Kazarovets, E.~V., Durlevich, O.~V., et al.\ 2017, Astronomy Reports, 61, 80. 
\bibitem[Searle \& Zinn(1978)]{1978ApJ...225..357S} Searle, L. \& Zinn, R.\ 1978, \apj, 225, 357. 
\bibitem[Sesar(2012)]{2012AJ....144..114S} Sesar, B.\ 2012, \aj, 144, 114.
\bibitem[Sesar et al.(2013)]{2013AJ....146...21S} Sesar, B., Ivezi{\'c}, {\v{Z}}., Stuart, J.~S., et al.\ 2013, \aj, 146, 21. 
\bibitem[Sesar et al.(2017a)]{2017AJ....153..204S} Sesar, B., Hernitschek, N., Mitrovi{\'c}, S., et al.\ 2017a, \aj, 153, 204. 
\bibitem[Sesar et al.(2017b)]{2017ApJ...844L...4S} Sesar, B., Hernitschek, N., Dierickx, M.~I.~P., et al.\ 2017b, \apjl, 844, L4. 
\bibitem[Shappee et al.(2014)]{2014ApJ...788...48S} Shappee, B.~J., Prieto, J.~L., Grupe, D., et al.\ 2014, \apj, 788, 48. 
\bibitem[Simion et al.(2019)]{2019MNRAS.482..921S} Simion, I.~T., Belokurov, V., \& Koposov, S.~E.\ 2019, \mnras, 482, 921. 
\bibitem[Simpson(2019)]{2019MNRAS.488..253S} Simpson, J.~D.\ 2019, \mnras, 488, 253. 
\bibitem[Skrutskie et al.(2006)]{2006AJ....131.1163S} Skrutskie, M.~F., Cutri, R.~M., Stiening, R., et al.\ 2006, \aj, 131, 1163. 
\bibitem[Smith et al.(2009)]{2009MNRAS.399.1223S} Smith, M.~C., Evans, N.~W., Belokurov, V., et al.\ 2009, \mnras, 399, 1223. 
\bibitem[Starkenburg et al.(2009)]{2009ApJ...698..567S} Starkenburg, E., Helmi, A., Morrison, H.~L., et al.\ 2009, \apj, 698, 567. 
\bibitem[Torrealba et al.(2015)]{2015MNRAS.446.2251T} Torrealba, G., Catelan, M., Drake, A.~J., et al.\ 2015, \mnras, 446, 2251. 
\bibitem[Vasiliev(2019)]{2019MNRAS.484.2832V} Vasiliev, E.\ 2019, \mnras, 484, 2832. 
\bibitem[Wang et al.(2021)]{2021MNRAS.504..199W} Wang, F., Zhang, H.-W., Huang, Y., et al.\ 2021, \mnras, 504, 199. 
\bibitem[Watkins et al.(2009)]{2009MNRAS.398.1757W} Watkins, L.~L., Evans, N.~W., Belokurov, V., et al.\ 2009, \mnras, 398, 1757.
\bibitem[White \& Rees(1978)]{1978MNRAS.183..341W} White, S.~D.~M. \& Rees, M.~J.\ 1978, \mnras, 183, 341.
\bibitem[Yang et al.(2019a)]{2019ApJ...880...65Y} Yang, C., Xue, X.-X., Li, J., et al.\ 2019a, \apj, 880, 65.
\bibitem[Yang et al.(2019b)]{2019ApJ...886..154Y} Yang, C., Xue, X.-X., Li, J., et al.\ 2019b, \apj, 886, 154. 
\bibitem[Yanny et al.(2000)]{2000ApJ...540..825Y} Yanny, B., Newberg, H.~J., Kent, S., et al.\ 2000, \apj, 540, 825. 
\bibitem[Yanny et al.(2009)]{2009AJ....137.4377Y} Yanny, B., Rockosi, C., Newberg, H.~J., et al.\ 2009, \aj, 137, 4377. 
\bibitem[York et al.(2000)]{2000AJ....120.1579Y} York, D.~G., Adelman, J., Anderson, J.~E., et al.\ 2000, \aj, 120, 1579. 
\bibitem[Yuan et al.(2020)]{2020ApJ...898L..37Y} Yuan, Z., Chang, J., Beers, T.~C., et al.\ 2020, \apjl, 898, L37. 
\bibitem[Zhao et al.(2012)]{2012RAA....12..723Z} Zhao, G., Zhao, Y.-H., Chu, Y.-Q., et al.\ 2012, Research in Astronomy and Astrophysics, 12, 723. 
\bibitem[Zinn et al.(2004)]{2004ASPC..327...92Z} Zinn, R., Vivas, A.~K., Gallart, C., et al.\ 2004, Satellites and Tidal Streams, 327, 92

\end{thebibliography}
\end{document}